%% file: main.tex
\documentclass[conference]{IEEEtran}
\IEEEoverridecommandlockouts
\usepackage{url}
\usepackage{cite}
\usepackage{booktabs} 
\usepackage{bigstrut}
\usepackage{amsmath}
\usepackage{listings}
\usepackage{lipsum}
\usepackage[utf8]{inputenc}
\usepackage[english]{babel} 
\usepackage[numbers]{natbib}
\usepackage{graphicx}
\usepackage{multirow}
\usepackage{subfig}
\usepackage{tabularx}
\usepackage{tikz}
\newcommand{\tikzcirclered}[2][red,fill=red]{\tikz[baseline=-0.5ex]\draw[#1,radius=#2] (0,0) circle ;}%
\newcommand{\tikzcirclegreen}[2][green,fill=green]{\tikz[baseline=-0.5ex]\draw[#1,radius=#2] (0,0) circle ;}%
\newcommand{\tikzcircleyellow}[2][yellow,fill=yellow]{\tikz[baseline=-0.5ex]\draw[#1,radius=#2] (0,0) circle ;}%
\usepackage{amsmath}
\usepackage{amssymb}
\usepackage{algorithm}
\usepackage[noend]{algpseudocode}
\usepackage{paralist}
\usepackage{color}
\usepackage{enumitem}
\usepackage{footnote}
\makesavenoteenv{tabular}
\makesavenoteenv{table}
\usepackage{stfloats}
\usepackage{courier}
\usepackage[T1]{fontenc}
\usepackage{float}
\usepackage{float}
\floatstyle{plaintop}
\restylefloat{table}
\usepackage{makecell}
\usepackage{amssymb}
\usepackage{pifont}
\usepackage{array}
\newcolumntype{P}[1]{>{\centering\arraybackslash}p{#1}}
\newcolumntype{L}{>{$}l<{$}}
\newcolumntype{C}{>{$}c<{$}}
\newcolumntype{R}{>{$}r<{$}}
\newcommand{\rom}[1]{\uppercase\expandafter{\romannumeral #1\relax}}
\usepackage{siunitx}

\definecolor{lightgray}{gray}{0.97}
\newcommand{\worst}[1]{{\bf \color{red}{#1}}}
\newcommand{\best}[1]{{\bf \textbf{#1}}}
\newcommand{\name}{SURFNet}

\newcommand{\Rey}{\mbox{\textit{Re}}}

\makeatletter
\newcommand{\ie}{i.e., }
\newcommand{\eg}{e.g., }

\newcommand{\linebreakand}{%
  \end{@IEEEauthorhalign}
  \hfill\mbox{}\par
  \mbox{}\hfill\begin{@IEEEauthorhalign}
}
\makeatother

\def\BibTeX{{\rm B\kern-.05em{\sc i\kern-.025em b}\kern-.08em
    T\kern-.1667em\lower.7ex\hbox{E}\kern-.125emX}}
\begin{document}

\title{SURFNet: Super-resolution of Turbulent Flows with Transfer Learning using Small Datasets\\
}

 \author{\IEEEauthorblockN{1\textsuperscript{st} Octavi Obiols-Sales}
 \IEEEauthorblockA{
 \textit{University of California, Irvine}\\
 Irvine, California \\
 oobiols@uci.edu}
 \and
 \IEEEauthorblockN{2\textsuperscript{nd} Abhinav Vishnu}
 \IEEEauthorblockA{
 \textit{Advanced Micro Devices, Inc.}\\
 Austin, Texas \\
 abhinav.vishnu@amd.com}
 \linebreakand
 \IEEEauthorblockN{3\textsuperscript{rd} Nicholas P. Malaya}
 \IEEEauthorblockA{
 \textit{Advanced Micro Devices, Inc.}\\
 Austin, Texas \\
 nicholas.malaya@amd.com}
 \and
 \IEEEauthorblockN{4\textsuperscript{th} Aparna Chandramowlishwaran}
 \IEEEauthorblockA{
 \textit{University of California, Irvine}\\
 Irvine, California \\
 amowli@uci.edu}
 }

\maketitle

\input{text/abstract}


\input{tables/TableQualitative}

\input{text/intro}

\input{text/background}

\input{text/methodology}

\input{text/experiment}
\input{text/results}

\input{text/related}
\input{text/conclusion}

\bibliographystyle{IEEEtranN}
\bibliography{./bib/octavi,./bib/distdl,./bib/FasterLearning,./bib/extra,./bib/vishnu,./bib/apoptosis,./bib/agd,./bib/mathstat}

\end{document}

%% file: text/abstract.tex
\begin{abstract}

Deep Learning (DL) algorithms are emerging as a key alternative to computationally expensive CFD simulations. 
However, state-of-the-art DL approaches require large \emph{and} high-resolution training data to learn accurate models. 
The size and availability of such datasets are a major limitation for the development of next-generation data-driven surrogate models for turbulent flows.  
This paper introduces \name, a transfer learning-based super-resolution flow network. 
\name\ primarily trains the DL model on {\em low-resolution} datasets and {\em transfer learns} the model on a handful of high-resolution flow problems -- accelerating the traditional numerical solver independent of the input size.
We propose two approaches to transfer learning for the task of super-resolution, namely one-shot and incremental learning.
Both approaches entail transfer learning on only \textbf{one} geometry to account for fine-grid flow fields requiring 15$\times$ less training data on high-resolution inputs compared to the tiny resolution ($64\times 256$) of the coarse model significantly, reducing the time for both data collection and training. 

We empirically evaluate \name 's performance by solving the Navier-Stokes equations in the turbulent regime on input resolutions up to \textbf{256$\times$} larger than the coarse model. 
On four test geometries and eight flow configurations unseen during training, we observe a consistent $2-2.1\times$ speedup over the OpenFOAM physics solver independent of the test geometry and the resolution size (up to $2048\times 2048$), demonstrating both resolution-invariance and generalization capabilities. 
Moreover, compared to the baseline model (aka oracle) that collects large training data at $256\times256$ and $512\times512$ grid resolutions, \name\ achieves the same performance gain while reducing the combined data collection and training time by $3.6\times$ and $10.2\times$, respectively. 
Our approach addresses the challenge of reconstructing high-resolution solutions from coarse grid models trained using low-resolution inputs (i.e., super-resolution) without loss of accuracy and requiring limited computational resources.  

\end{abstract}

%% file: tables/TableQualitative.tex
\newcommand*\TableQualitative{
	\begin{table*}
		\scriptsize
	\begin{center}
\setlength{\tabcolsep}{4.5pt}
\renewcommand{\arraystretch}{0.7}
		\begin{tabular}{c c c c c c c c c c}
			
			\toprule
			
			Related Work 
			& \makecell[l]{Target\\Flow (PDE)} 
			& \makecell[l]{Test\\ Geometry \\ Unseen \\ in Training} 
			& \makecell[l]{Meets\\convergence\\constraints} 
			& \makecell[l]{Resolution\\Invariant} 
			& \makecell[l]{Training\\data\\collected\\on\\coarse-grids} 
			& \makecell[l]{Highest\\Test\\Spatial\\Resolution} 
			& \makecell[l]{Error\\metric} 
			& \makecell[l]{Error \\value\\ (best)} 
			& \makecell[l]{Technique} \\ [0.2ex]
			
			\midrule

			\makecell[l]{Fourier Neural Operator~\cite{fourieroperator}}    
			& \makecell[l]{Darcy}    
			& \tikzcirclered{1.8pt}      
			& \tikzcirclered{1.8pt}      
			& \tikzcirclegreen{1.8pt}    
			& \tikzcirclered{1.8pt}    
			& \makecell[l]{$421\times421$}
			& \makecell[l]{RE}
			& \makecell[l]{\num{1e-2}}
			& \makecell[l]{NO} \\

			\makecell[l]{Fourier Neural Operator~\cite{fourieroperator}}    
			& \makecell[l]{Navier-Stokes}    
			& \tikzcirclered{1.8pt} 
			& \tikzcirclered{1.8pt} 
			& \tikzcirclegreen{1.8pt} 
			& \tikzcirclered{1.8pt}  
			& \makecell[l]{$256\times256$}
			& \makecell[l]{RE}
			& \makecell[l]{\num{8.6e-3}}
			& \makecell[l]{NO} \\

			\makecell[l]{Graph Kernel Network~\cite{graphkernelnetwork}}     
			& \makecell[l]{Darcy}     
			& \tikzcirclered{1.8pt}  
			& \tikzcirclered{1.8pt} 
			& \tikzcirclegreen{1.8pt}  
			& \tikzcirclered{1.8pt}   
			& \makecell[l]{$241\times241$}		 
			& \makecell[l]{Relative $L_2$} 
			& \makecell[l]{\num{3.7e-2}}
			& \makecell[l]{NO}\\

			\makecell[l]{\citeauthor{bhattacharya}~\cite{bhattacharya}}     
			& \makecell[l]{Darcy}     
			& \tikzcirclered{1.8pt}  
			& \tikzcirclered{1.8pt} 
			& \tikzcirclegreen{1.8pt}  
			& \tikzcirclered{1.8pt}   
			& \makecell[l]{$421\times421$}
			& \makecell[l]{RE}
			& \makecell[l]{\num{2e-3}}
			& \makecell[l]{NO}\\

			\makecell[l]{MeshFreeFlowNet~\cite{meshfreeflownet}}     
			& \makecell[l]{Rayleigh-Bénard}     
			& \tikzcirclered{1.8pt}  
			& \tikzcirclered{1.8pt} 
			& \tikzcirclegreen{1.8pt}  
			& \tikzcirclered{1.8pt}   
			& \makecell[l]{$4\times512$}
			& \makecell[l]{NMAE}
			& \makecell[l]{\num{3.3e-3}}
			& \makecell[l]{CNN}\\
			
			\makecell[l]{\citeauthor{denoisingsuperresolution}~\cite{denoisingsuperresolution} }
			& \makecell[l]{Laminar}     
			& \tikzcirclered{1.8pt}  
			& \tikzcirclered{1.8pt} 
			& \tikzcirclegreen{1.8pt}  
			& \tikzcirclered{1.8pt}   
			& \makecell[l]{$200\times200$}
			& \makecell[l]{RE}
			& \makecell[l]{0.025} 
			& \makecell[l]{CNN}\\
			
			\makecell[l]{TFNet~\cite{TFNet}}
			& \makecell[l]{Rayleigh-Bénard}     
			& \tikzcirclered{1.8pt}  
			& \tikzcirclered{1.8pt} 
			& \tikzcirclered{1.8pt}  
			& \tikzcirclered{1.8pt}   
			& \makecell[l]{$1792\times256$}
			& \makecell[l]{RMSE}
			& \makecell[l]{200} 
			& \makecell[l]{CNN}\\

			\makecell[l]{\citeauthor{autodesk}~\cite{autodesk}}     
			& \makecell[l]{Laminar}     
			& \tikzcirclegreen{1.8pt}  
			& \tikzcirclered{1.8pt} 
			& \tikzcirclered{1.8pt}  
			& \tikzcirclegreen{1.8pt}   
			& \makecell[l]{$128\times256$}
			& \makecell[l]{RME}
			& \makecell[l]{1.76\%} 
			& \makecell[l]{CNN}\\
			
			\makecell[l]{CFDNet~\cite{cfdnet}}     
			& \makecell[l]{Navier-Stokes}     
			& \tikzcirclegreen{1.8pt} 
			& \tikzcirclegreen{1.8pt} 
			& \tikzcirclered{1.8pt} 
			& \tikzcirclegreen{1.8pt} 
			& \makecell[l]{$64\times256$}
			& \makecell[l]{RME}
			& \makecell[l]{0\%} 
			& \makecell[l]{CNN}\\
			
			\makecell[l]{Smart-fluidnet~\cite{SC19}}     
			& \makecell[l]{Eulerian}     
			& \tikzcirclegreen{1.8pt}  
			& \tikzcircleyellow{1.8pt} 
			& \tikzcirclegreen{1.8pt}  
			& \tikzcircleyellow{1.8pt}   
			& \makecell[l]{$1024\times1024$}
			& \makecell[l]{MAE} 
			& \makecell[l]{\num{9e-3} }
			& \makecell[l]{CNN}\\

			\midrule

			\makecell[l]{\textbf{\name\ (this paper)}}     
			&\makecell[l]{Navier-Stokes}     
			& \tikzcirclegreen{1.8pt}  
			& \tikzcirclegreen{1.8pt} 
			& \tikzcirclegreen{1.8pt}  
			& \tikzcirclegreen{1.8pt}   
			& \makecell[l]{$2048\times2048$		 }
			& \makecell[l]{RME}
			& \makecell[l]{0\%} 
			& \makecell[l]{TL}\\

			\bottomrule
		\end{tabular}
		\caption{\small Comparing state-of-the-art approaches in DL for 2D CFD simulations on nine different features. \name\ is a novel network-based transfer learning (TL) framework that (1) generates accurate solutions up to $2048\times2048$ spatial resolutions for turbulent flows from low-resolution models, (2) generalizes to unseen-in-training geometries, (3) meets the original convergence constraints of traditional CFD solvers, and (4) collects data at low-resolution on coarse grids for training -- whereas prior works target non-turbulent or non-viscous flows~\cite{graphkernelnetwork,bhattacharya,SC19,autodesk}, only test on geometry domains that were part of the training phase~\cite{meshfreeflownet,fourieroperator,TFNet}, replace partly~\cite{SC19} (left yellow dot) or entirely~\cite{autodesk,fourieroperator,meshfreeflownet,TFNet} traditional solvers with a neural network surrogate not meeting convergence constraints, and most importantly, train with data downsampled from high-resolution simulations or are unsupervised (right yellow dot)~\cite{meshfreeflownet,fourieroperator,bhattacharya,graphkernelnetwork,SC19}. \emph{NO} stands for Neural Operator and \emph{CNN} for Convolutional Neural Network. \label{tab:intro}}
	\vspace{-2em}
	\end{center}
	\end{table*}

}

%% file: text/intro.tex
\vspace{-2em}
\section{Introduction}
\label{sec:introduction}
\begin{figure}[htbp]
\small
	\centerline{\includegraphics[scale=0.6]{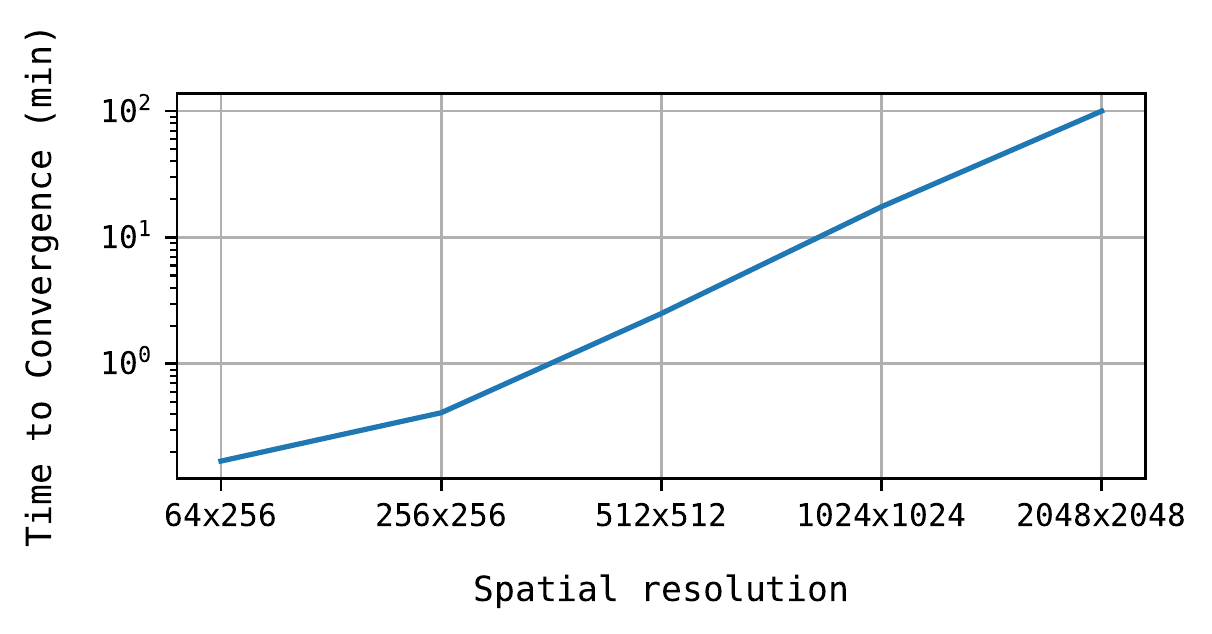}}
\caption{\small Time to convergence for flow around a NACA4415 airfoil at
	different spatial resolutions on a dual-socket Intel Xeon Gold 6148 CPU (total 40 cores).\vspace{-1em}}
\label{fig:collectiontime}
\end{figure}

Computational Fluid Dynamics (CFD) simulations that solve the complex Navier-Stokes equations are ubiquitous in both laminar and turbulent flows~\cite{malaya,casacuberta,blazek,openfoam,RANS,behnam,Mostafazadeh:aa}. Engineering systems of interest such as aerospace design exploration (\eg designing airfoils for aircraft wings) require resolving fine-scale physics in the turbulent regime to produce high-fidelity solutions.
To account for more aspects of the physical phenomena being modeled necessitates an increase in the resolution of the system, resulting in high computational costs. 
Figure~\ref{fig:collectiontime} shows the time-to-solution with increasing grid size for a turbulent flow simulation around a National Advisory Committee for Aeronautics (NACA) airfoil. 
A dual-socket, 40-core system requires ten seconds at a resolution of $64\times256$, and 100 minutes at $2048\times2048$ for a single airfoil shape in one flow configuration (one test case). 
In aircraft design, there are typically many airfoil shapes~\cite{naca-airfoils}, each of which requires simulations performed across a range of parameters such as Reynolds (Re) numbers, various angles of pitch, yaw, and roll to account for rotation as well as different angles of attack, resulting in thousands of test cases.
To address this challenge, \emph{super-resolution} is an approach to reconstruct fine-scale flow physics from coarse-grid solutions.

Although mathematical models such as Large Eddy Simulations (LES)~\cite{LES} and Reynolds-Averaged Navier-Stokes (RANS)~\cite{RANS} that relate coarse-scale physical effects with fine-scale solutions are common place, most recent applications of super-resolution employ deep learning (DL) \cite{STORM,singleimageSR,DLforSR,meshfreeflownet,fourieroperator,denoisingsuperresolution,graphkernelnetwork,bhattacharya}.
This popularity of DL is due to its tremendous success in natural language processing~\cite{SOTANLP} and computer vision~\cite{imagenetinminutes,43022}.

\TableQualitative

Table~\ref{tab:intro} summarizes the state-of-the-art approaches for accelerating CFD simulations using DL and categorizes them across several features. 
First, the majority of Convolutional Neural Network (CNN) based approaches are finite-dimensional maps and hence lack resolution-invariance~\cite{autodesk,cfdnet,unet,zhu,TFNet}.
Mesh-free, resolution-invariant DL methods are a potential alternative to query fine-scale flow physics~\cite{fourieroperator,graphkernelnetwork,bhattacharya}.
However, these approaches use low-resolution data downsampled from high-resolution solutions to train the network.
As a result, current mesh-free methods suffer from the same limitation as classical CNN-based approaches that require a large number of computationally expensive simulations to collect training data at high resolutions. 
Practical CFD simulations require high spatial resolutions (such as $1024\times 1024$) according to NASA's FUN3D and CFL3D solvers~\cite{propergrid} making current DL approaches computationally prohibitive (see Figure 1). 
Second, most approaches~\cite{meshfreeflownet,fourieroperator,graphkernelnetwork,bhattacharya,denoisingsuperresolution,TFNet} lack generalization to unseen geometries where the test geometry is a subset of the training data -- a key limitation for CFD researchers/practitioners. 
Third, only a limited number of approaches support super-resolution of turbulent flows that are significantly more challenging than laminar flows~\cite{denoisingsuperresolution} and ubiquitous with a wide range of applications in aerodynamics, atmospheric science, turbomachinery, and propulsion~\cite{cfdnet,zhu,unet,meshfreeflownet}. 
Lastly, it's common to use DL algorithms as a pure surrogate model that entirely replaces the CFD solver, thereby not providing the same convergence guarantees as the latter. 
There is a pressing need to design DL based models for super-resolution
of turbulent flows that (a) eliminates the need for extensive data collection at high resolutions, (b) provides discretization and resolution-independent acceleration, and
(c) can generalize to unseen geometries and flow configurations while meeting
the convergence guarantees of the traditional physics solvers.

To address the above need, we develop \textbf{\name\ } (\textbf{SU}per-\textbf{R}esolution \textbf{F}low \textbf{Net}work), a novel approach to reconstruct fine-scale flow physics from coarse grid data by primarily training the DL model on {\em low-resolution} inputs.
Using this {\em coarse-model} (CM), \name\ {\em transfer learns} the model on high-resolution turbulent flow solutions, significantly reducing the overall data collection time and the total size of the training set. 
More specifically, this paper makes the following contributions:

\begin{itemize}

\item First, we present an 8-layer CNN that is adequately expressive to capture different geometries and flow configurations. The CNN is trained on {\em low-resolution} inputs ($64\times256$ grid
	size) using ten geometries and nine variations of each geometry to produce a {\em coarse-model} (CM) (Section~\ref{subsec:CM}). 

\item Then, to enable efficient super-resolution, we propose two variations of transfer learning -- 1) one-shot transfer learning (OSTL): learning is conducted from CM to the target resolution in a single shot, and 2) incremental transfer learning (ITL): learning is done step-by-step incrementally on the training set of intermediate discretizations up to the target resolution (Section~\ref{subsec:methodology:surfnet}). 
We show that one-shot learning is inadequate for reaching oracle (\ie a full model trained at high resolutions) accuracy, and incremental learning where the model is fine-tuned with data from intermediate discretizations is critical for reaching best-case speedups, especially at higher target discretizations (Section~\ref{subsec:results:loss}).

\item We empirically evaluate \name\ by solving the RANS equations in the turbulent regime on four geometries and eight flow configurations unseen during training. 
\name\ achieves a consistent speedup of 2 - 2.1$\times$ with ITL at up to $2048\times2048$ spatial resolutions over the OpenFOAM physics solver independent of the resolution size and test geometry, demonstrating both resolution-invariance and generalization capabilities (Section~\ref{subsec:results:performance}).

\item \name\ eliminates the need to collect exhaustive training datasets at high resolutions to account for fine-scale physical phenomena. 
This computational efficiency enables \name\ to achieve oracle accuracies while significantly reducing the size of the training dataset by $15\times$, consequently reducing the combined data collection and training time by $3.6\times$ and $10.2\times$, respectively at $256\times256$ and $512\times512$ grid sizes (Section~\ref{subsec:results:baseline}).

\end{itemize}

%% file: text/background.tex
\section{Background}
\label{sec:background}

We use the steady incompressible RANS equations to solve turbulent flows. The RANS governing equations describe the fluid motion as follows: 

\vspace{-1em}
\begin{align}
  \frac{\partial \bar{U}_i}{\partial x_i} &= 0\label{eq:continuity}\\ \bar{U}_j
  \frac{\partial \bar{U}_i}{\partial x_j} &= \frac{\partial}{\partial
  x_j}\left[ -\left(\bar{p}\right)\delta_{ij}
  +\left(\nu+\nu_t\right)\left(\frac{\partial \bar{U}_i}{\partial x_j}
  + \frac{\partial \bar{U}_j}{\partial x_i}\right)\right]\label{eq:momentum}
\end{align} 

where $\bar{U}$ is the mean velocity,
$\bar{p}$ is the kinematic mean pressure, $\nu$ is the fluid
viscosity, and $\nu_t$ is the eddy viscosity.

The RANS equations provide a time-averaged solution to the incompressible Navier-Stokes equations at the expense of yielding a non-closed equation. Closing the equation is usually done through Boussinesq's
approximation~\cite{boussinesq}, which yields the eddy viscosity. The eddy viscosity is found through turbulence modeling. The Spalart-Allmaras one-equation model shown below provides
a transport equation to compute the modified eddy viscosity, $\tilde{\nu}$.

\begin{multline}
\bar{U}_i \frac{\partial \tilde{\nu}}{\partial x_i } = C_{b1} \left(1-f_{t2}\right) \tilde{S} \tilde{\nu} - \left[C_{w1}f_w - \frac{C_{b1}}{\kappa^2} f_{t2} \right] \left(\frac{\tilde{\nu}}{d}^2\right) \\ + \frac{1}{\sigma} \left[\frac{\partial}{\partial x_i} \left( \left(\nu+\tilde{\nu} \right) \frac{\partial \tilde{\nu}}{\partial x_i}\right) +C_{b2} \frac{\partial \tilde{\nu}}{\partial x_j}\frac{\partial \tilde{\nu}}{\partial x_j} \right]
\label{eq:SA}
\end{multline}

Then, we can compute the eddy viscosity from $\tilde{\nu}$ as $\nu_t = \tilde{\nu} f_{v1}$. These equations represent the most popular implementation of the Spalart-Allmaras model.
The terms $f_{v1}$, $\tilde{S}$, and $f_{t2}$ are specific to the model. They contain, for example, first order flow features (magnitude of the vorticity). $C_{b1}$, $C_{w1}$, $C_{b2}$, $\kappa$, and $\sigma$ are constants also specific to the model, found experimentally. $d$ is the closest distance to a solid surface. The first original reference of the model details the equations and the values of these constants~\cite{SpalartAllmaras}.

%% file: text/methodology.tex
\section{SURFN\lowercase{et}: Transfer Learning for Super-Resolution Flow Simulations}
\label{sec:methodology}

Our objective is to accelerate the convergence of high-resolution turbulent flow simulations.
We achieve this by reconstructing fine-grid flow solutions from coarse grid models referred to as \emph{super-resolution} with minimum data collection at higher resolutions. 

In this section, we first describe the strategy used to accelerate the convergence of the equations described in Section~\ref{sec:background} with DL. Then, we describe the design of the CNN architecture to train the \emph{coarse model} (CM) with low-resolution data.
Finally, we present \name, a novel transfer learning-based super-resolution approach that relaxes the demand for generating expensive and time-consuming training data to train accurate models for discretizations at higher resolutions.

\subsection{Accelerating the convergence of fluid simulations}

To accelerate the numerical convergence of the equations in Section~\ref{sec:background}, we define a finite-dimensional parametric map $G$ such that 
$x^{N} = G\left(x^{I};\theta_{wl}\right)$ where $x^{N}, x^{I}\in \mathbb{R}^{m \times
n \times z}$ represent the tensors of $z$ flow variables on a $m\times n$ grid domain at convergence and any intermediate iteration respectively, \emph{N} is the number
of iterations required for convergence with the physics solver, and \emph{I} $(0\leq I \leq N-1)$
is any intermediate iteration of the solver. 

At each $I$, the physics solver computes four flow fields in the 2D grid domain of the simulation (\ie $z=4$): the x-velocity component ($U$), the y-velocity component ($V$), the pressure ($P$), and the modified eddy viscosity ($\nu$). 
If the grid resolution is $m\times n$, each field is of size $m\times n$. 
We concatenate the four flow variables together, generating an input tensor, $x^{I}$ of size $m \times n \times 4$. The output tensor has the same shape as the input tensor. 
The difference between the input and the output is that the latter has the flow field values at steady-state, $x^{N}$. 
As a result, $G$ parameterizes the solution operator that takes an intermediate flow field and maps it to its steady-state solution.

A popular candidate for learning $G$ is CNNs. 
However, as seen from Table~\ref{tab:intro}, the majority of the CNN-based approaches present a solution that does not satisfy the conservation laws~\cite{autodesk,unet,cnnfoil,spatialconvolution}, which is sub-optimal for CFD practitioners. 
There have been recent attempts to provide physically consistent solutions.
However, these methods either severely restrict the generalization capacity of the network to specific flow configurations~\cite{meshfreeflownet,SC19} or present boundary-constrained loss functions that require training a new model for every new instance of the same fluid problem~\cite{pinns,deepxde,sciann,NSFnets}.

To circumvent these limitations, other works~\cite{cfdnet,maulik} train a domain-agnostic DL model and subsequently augment the model's inference with a \textit{refinement} stage, where the physics solver constrains the model's prediction to reach accelerated convergence. Refinement consists of (a) feeding the CNN's prediction back into the physics solver, (b) imposing the boundary conditions, and (c) running the solver to convergence. 
We adhere to the latter approach because it allows us to retain the advantages of domain-agnostic training while simultaneously meeting the original convergence constraints of the physics solver.
The solution, therefore, has a 0\% relative mean error. 

\subsection{Coarse Grid Network} 
\label{subsec:CM}

\begin{figure*}[htbp]
\centering
\includegraphics[width=0.92\textwidth]{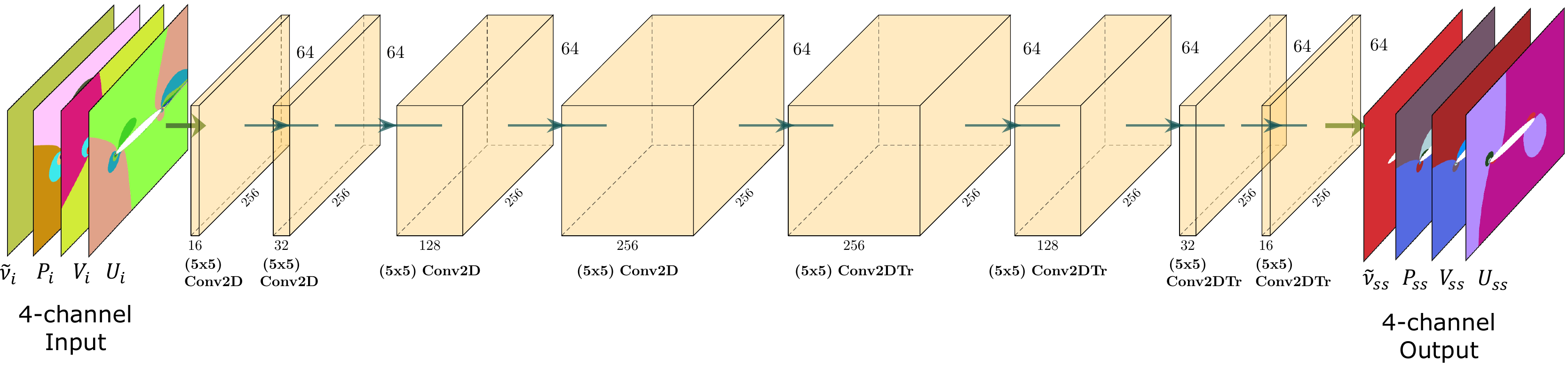}
	\vspace{-0.7em}
	\caption{\small CNN and its input-output representation. The CNN is a symmetric, fully convolutional-deconvolutional neural network. The input and the output are a 4-channel tensor image: each channel represents one flow variable (x-velocity $U$, y-velocity $V$, pressure $P$ and modified eddy viscosity $\tilde{\nu}$). The difference between the input and the output tensors is that the former has the flow values of an intermediate iteration (i), while the latter is the flow field at steady-state (ss).}
\label{fig:network}
\end{figure*}

We aim to learn a generalizable model that can predict flow around rotated geometries (\ie varying pitch angle, $\theta$) and changing flow fields (\ie angle of attack, $\alpha$) as they are critical features in design exploration and shape optimization. 
For instance, it is common practice to simulate the flow field by varying the angle of the incoming flow and rotating the airfoil geometry~\cite{spatialconvolution}.

To capture such complex flow phenomena, we design an eight-layer, convolutional-deconvolutional neural network (4 convolution layers followed by 4 deconvolution layers). The number of filters are, in order: $16$, $32$, $128$, $256$, and $256$, $128$, $32$, $16$. An illustration can be seen in Figure~\ref{fig:network}. The choice of the architcture is motivated by two reasons. First, recent works have successfully leveraged similar designs for physical system emulation~\cite{autodesk,cfdnet}. Second, the convolution operator is an optimal candidate to extract existing spatial correlations in and among the fluid variables. We use a deeper network compared to other approaches~\cite{autodesk,cfdnet} (8- vs 6-layer CNN) because the training dataset is larger and has more features to learn (\eg rotation). All layers have a filter size of $5\times 5$ and a stride of $1$. This overlap captures both the short and long (spatial) range dependencies between the flow variables while covering all regions of the flow field present in the input image. We use the LeakyReLU activation function for all layers because the output image contains real-valued variables.

\subsection{Why Transfer Learning?}
\label{subsec:methodology:TL}
The CNN in Figure~\ref{fig:network} relies on intermediate and final solutions of flow simulations to train the solution operator, $G$. 
The ideal scenario in machine learning is the availability of vast amounts of labeled training data for supervised models.
However, both collection of large-scale high-resolution data and training with large input sizes are prohibitively expensive (see Figure~\ref{fig:collectiontime}). 
An alternate approach is to use the convolutional operator calibrated with low-resolution data to predict high-resolution solutions.
Although this is feasible, the accuracy reduces dramatically, especially for turbulent flows as we show empirically in Section~\ref{sec:results}. 
Therefore, CNN-based approaches demand end-to-end re-training for different resolutions and discretizations to achieve constant error.

The above limitations motivate the use of \emph{transfer learning} (TL) to both solve the problem of insufficient training data at higher resolutions and to achieve resolution-invariance across discretizations. 
TL is a popular technique that relies on transferring a trained
model across different learning tasks or from one model to another~\cite{tan2018survey, zhuang2020comprehensive}.
It has been successfully applied to different applications such as drug discovery~\cite{TLdrug}, disease detection~\cite{TLCOVID}, natural language processing~\cite{TLNLP}, and machine fault diagnosis~\cite{TLmachine}.

We propose another novel application of TL that leverages the correlations inherent between low- and high-resolution inputs of turbulent flows.
Low-resolution solutions of fluid simulations contain critical information that can be effectively extracted for high-resolution flow prediction: low-resolution
solutions capture all present flow structures~\cite{casacuberta}, even complex ones (for instance, the wake region behind solid bodies after flow separation).
However, they do not resolve them, as high-resolution grids do.
Therefore, given a pre-trained CNN model on large datasets at low-resolution
(\ie \textit{coarse model}), our objective is to transfer the trained model to the target fine-grid discretization to append the unresolved flow information.
An advantage of our super-resolution TL methodology is that the input is \emph{truly} from coarse-grid simulations, not downsampled from high-resolution data as in the state-of-the-art~\cite{fourieroperator,graphkernelnetwork,bhattacharya}\footnote{Data downsampled from higher resolutions is not the same as data generated from coarse-grid solutions. Low-resolution data downsampled from high-resolution solutions contain information of resolved complex flow structures, whereas this information is inexistent in coarse-grid solutions.}.
This eliminates the requirement for generating computationally demanding (and in some cases intractable) data to build a high-resolution model.
In the rest of this section, we describe the transfer learning pipeline and how we account for fine-scale physics and dynamics. 

\subsection{Super-Resolution with Transfer Learning}
\label{subsec:methodology:surfnet}
\begin{figure*}[htbp]
\centering
\includegraphics[width=0.95\textwidth]{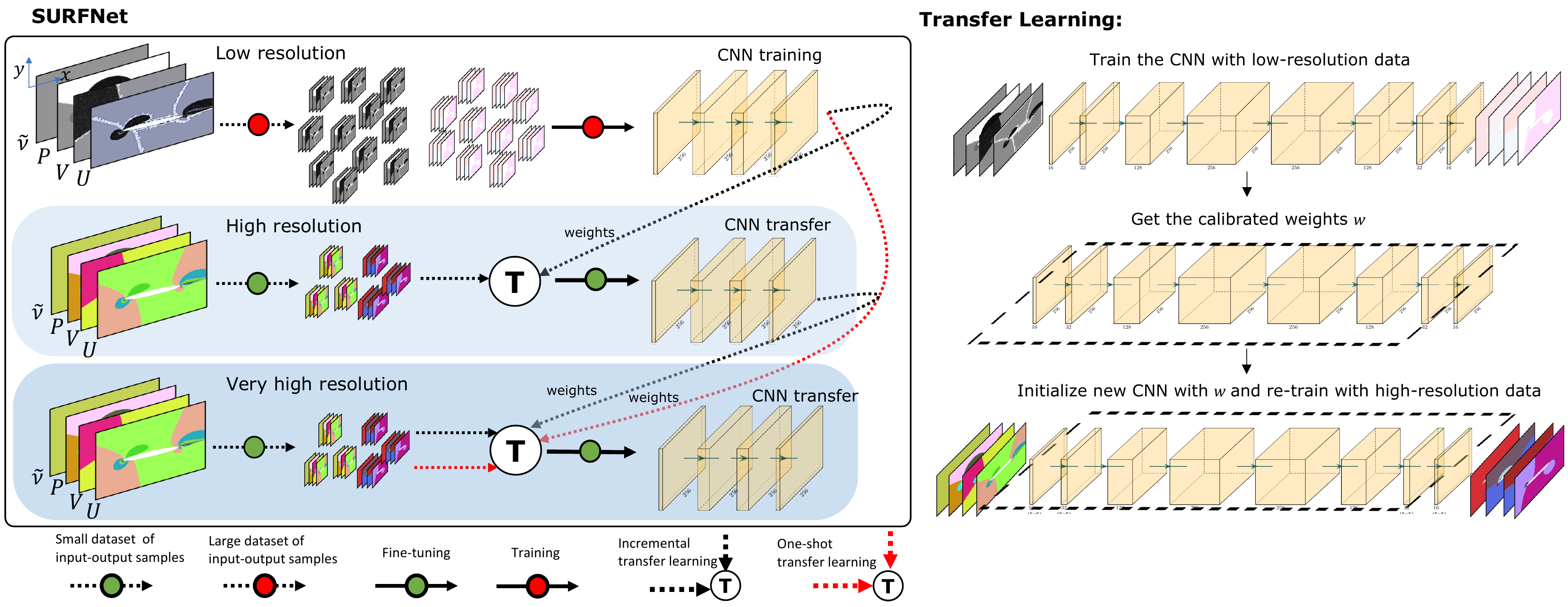}
	\vspace{-0.75em}
	\caption{\small SURFNet (left) and its inductive transfer learning (right)
  for super-resolution of turbulent flows. A large dataset is collected at
  low resolutions to train the CNN and obtain the coarse model. Small data is
	collected at higher resolutions and SURFNet transfers the model weights from the coarse model to train the fine-scale model using either one-shot (red) or incremental transfer learning (black).} 
\label{fig:transfer}
\end{figure*}

Given $r_c: x_c \times y_c$ where $r_c$ is the coarse-grid resolution corresponding to the discretization $x_c \times y_c$, the super-resolution task of \name\ is to recover solutions at fine-scales $r_f: x_f \times y_f \, | \, f = 1, ..., t$ where $x_f \times y_f$ is a fine-grid discretization and $r_f > r_c, \, \forall f$. 
We implement network-based transfer learning as illustrated in Figure~\ref{fig:transfer} and propose two variations for super-resolution as described below.

\emph{\textbf{One-Shot Transfer Learning (OSTL)}}. In this approach, model
weights are transferred from the CM trained with $r_c$ to the target resolution, $r_t$ in a single training step. 
For example, if we desire to recover high-resolution flow fields at $2048\times2048$, we transfer learn from $64\times256$ to $2048\times 2048$ in a  \emph{single shot}. 
This approach is illustrated in Figure~\ref{fig:transfer} with a red, dashed line.

\emph{\textbf{Incremental Transfer Learning (ITL)}}. An alternative approach to
OSTL is to train the CNN with the large-scale low-resolution dataset and
perform TL in a step-wise manner.
That is, instead of transfer learning from $r_c$ to the target resolution directly,
$r_t$, we pass through intermediate resolutions step-by-step from the low
resolution to the target high resolution (\ie $r_c$->$r_1$->$r_2$...->$r_t$). 
For example, if we desire to recover high-resolution flow fields at $512\times 512$, we transfer learn from $64\times256$ to $256\times 256$, and finally from $256\times 256$ to $512\times512$. 
A step size of 1 allows the model to incorporate new information from each intermediate discretization.
We further discuss the impact of step size on the predictive accuracy in Section~\ref{sec:results}.
The black dashed line in Figure~\ref{fig:transfer} shows this variation of TL.
In ITL, the model learns from more data than OSTL without overfitting while still reducing the overall data collection since we avoid heavy data collection at any specific high resolution.

To recover the solution at the desired target discretization, we perform the steps outlined below for both transfer learning approaches (\ie OSTL and ITL).

1. \emph{Coarse-grid data collection.} The training dataset is created by performing large-scale simulations at low-resolution discretizations, $x_c \times y_c$ using the physics solver detailed in Section~\ref{sec:openfoam}.

2. \emph{Training.} After generating low-resolution training data, the CNN in Figure~\ref{fig:network} is trained with this large dataset. By carefully controlling for under- and over-fitting of the network with a validation dataset, we obtain the coarse model. 

3. \emph{Fine-grid data collection.} Due to the challenge of data acquisition at fine-scale discretizations, $x_f \times y_f$, we limit the number of simulations at high resolutions to the bare minimum to create the \emph{transfer dataset}. 
In Section~\ref{sec:results}, we show that a single geometry is sufficient to transfer the coarse-grid features and recover the fine-scale physics.

4. \emph{Transfer learning.} In this paper, we choose to implement inductive transfer learning, where we reuse the network (including its structure and parameters) that was pre-trained to learn the source model (\ie CM)~\cite{TLresnet}. 
Since the source and target domains are the same, we intuitively expect this technique to preserve the common features extracted between the two learning tasks. 
We treat the coarse model as a feature extractor of high-resolution flow fields. 
All layers of the CNN use a stride of 1 to not reduce the dimensionality of the input-to-output map during the low-resolution training phase. 
However, for high-resolution input-output pairs, this model \emph{is} a low-dimensional representation -- an extractor of the prevalent flow features across discretizations (\eg flow effects due to the boundary conditions, fields' shape in the free stream, and flow variations due to the change in $\alpha$ and $\theta$). 
Since coarse discretizations do not have enough domain points to define accurate flow field solutions in areas of strong gradients, the features extracted need to be fine-tuned but not re-learned. 
The transferred network is updated by fine-tuning the weights as follows (see Figure~\ref{fig:transfer}).
\vspace{-0.4em}
\begin{enumerate}
	\item[a.] Re-initialize the transferred network with the weights obtained from the CM for OSTL.  
For ITL, the transferred network is initialized with the weights from the largest model pre-trained at resolution $r_f$ such that $r_f < r_t$.

\item[b.] Start training the transferred network with the transfer dataset. 
At this stage, it is critical to append the fine-grid flow features. 
Since we start with a good initial calibration of the weights from a pre-trained model of the same domain, only \emph{fine-tuning} is required to update the weights of the transferred network. 
Fine-tuning of network parameters is done with a low learning rate (\ie small updates to the weights) and for very few epochs (1 or 2 at most) to avoid overfitting to the geometry (or geometries) of the small transfer dataset while preserving the generalization capacity of the model. 
\end{enumerate}
\vspace{-0.4em}
In summary, SURFNet is a TL-based super-resolution flow network that learns three distinct CNN models to reconstruct high-resolution turbulent flow fields -- (1) the low-resolution CM, (2) the OSTL model, and (3) the ITL model.

%% file: text/experiment.tex
\section{Experiment Setup}
\label{sec:experiment}
In this section, we first describe the case study to evaluate \name's potential for super-resolution.
Then, we describe the low- and high-resolution datasets for training, transfer learning, validation, and testing and outline the training process of the CM.

\vspace{-0.5em}
\subsection{Case Study}
\label{sec:casestudy}
We consider external aerodynamics as the paradigm for aerospace design space exploration. 
Understanding flow around solid bodies is an important research topic for industrial aerospace applications, and airfoils are the core geometries in aerodynamics studies. 
In real scenarios, the exploration involves different geometries (for instance, airfoil shapes) simulated under various flow configurations such as rotation of the solid body and wind angle. 
Therefore, to apply to a large class of CFD problems, challenges in generalizing to unseen geometries need to be addressed.
In this paper, we aim to evaluate \name's ability to accelerate the flow around solid bodies such as airfoils and cylinder. 
Accordingly, these geometries are excluded from the training and transfer learning datasets. 
We resolve turbulent flow around solid bodies at a Reynolds number of \num{6e5} using the equations presented in Section~\ref{sec:background}. This setup - the flow regime (turbulent flow), the geometries (extrapolating to NACA airfoils), and the grid resolutions - is representative of real NASA case studies~\cite{propergrid}.

\subsection{Dataset Creation}
\label{sec:datasets}
We create a total of 15 distinct datasets and perform the simulations of all flow configurations with the solver described in Section~\ref{sec:openfoam}.
Table~\ref{tab:datasets} summarizes the datasets, including the composition, resolution, and scope of each dataset.

\input{tables/TableMeshDataset}

\TableMeshDataset

\begin{figure}[htbp]
\centering
\small
\includegraphics[width=0.95\columnwidth]{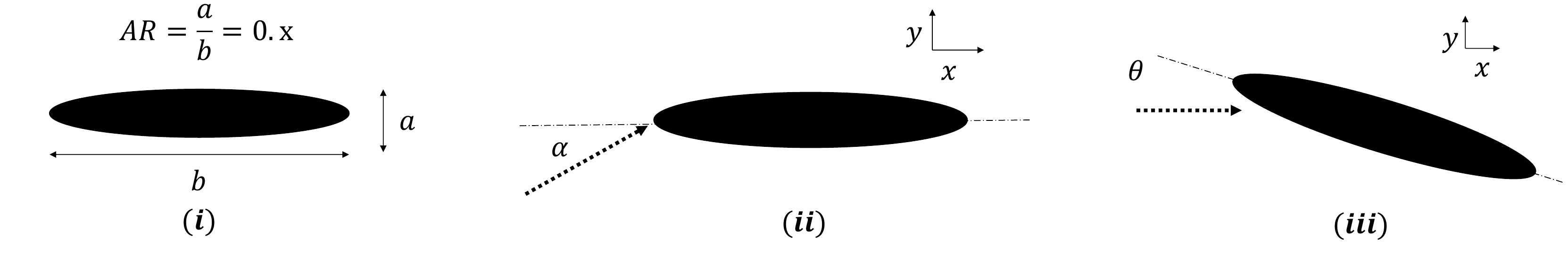}
	\caption{\small Geometry configurations used in training. The dotted arrow shows the direction of the incoming flow and the dashed line shows the chord of the solid body. Sketch of the (i) ellipse aspect ratio, (ii) angle of attack $\alpha$, and (iii) pitch angle $\theta$.}
	
\label{fig:ellipse_pitch_attack}
\end{figure}

\emph{\textbf{Training dataset}}. First, we collect a large-scale, low-resolution dataset to train the coarse model (CM), at a $64\times 256$ resolution (a common resolution for low-resolution solutions of similar case studies~\cite{propergrid}).
The core of this dataset is formed by simulations of flow around 10 different ellipses obtained by changing the aspect ratio $AR$ that is defined as the ratio of the vertical to the horizontal semiaxis length, as shown in Figure~\ref{fig:ellipse_pitch_attack}. 
The $AR$'s considered for the training dataset are: $0.05$, $0.07$, $0.09$, $0.1$, $0.15$, $0.2$, $0.25$, $0.35$, $0.55$, and $0.75$. 
For each ellipse, we consider 9 flow variations: five different angles of attack, $\alpha$, and four different pitch angles, $\theta$, chosen randomly for each ellipse in the range between $-2 - 6^{\circ}$ for a total of 90 flow configurations. We choose from thin to thick ellipses and angles of attack to cover a wide spectrum of physical phenomena that is commonplace in aerospace design exploration. For instance, the angle of attack is an important variable in determining the magnitude of the force of lift. 
The $\alpha$ angles are obtained by changing the direction of the flow while maintaining the angle between the chord of the solid body and the cartesian x-direction at $0^{\circ}$. 
The $\theta$ angles are obtained by pitching the nose of the solid body up or down and maintaining a flow direction at a 0-degree angle with its longitudinal axis.
These configurations are illustrated in Figure~\ref{fig:ellipse_pitch_attack}. 

\emph{\textbf{Transfer dataset}}. Next, we collect a small-scale, high-resolution dataset. 
The transfer dataset consists of flow around \emph{one} unique geometry, ellipse $AR=0.1$. 
For this geometry, we consider 6 flow variants: three $\alpha$ angles and three $\theta$ angles, chosen randomly as in the training dataset. 
Note that we collect a transfer dataset at each target resolution: $256\times 256$, $512\times 512$, $1024\times 1024$, and $2048\times2048$. 
Each of the four transfer datasets consists of the same flow configurations -- the only difference between them is the grid resolution. 
This choice is notable for two reasons. 
First, it stresses the extreme case of whether a single \emph{and} identical geometry is sufficient to recover high-resolution flow fields at any target resolution, both with OSTL and ITL. 
Second, each transfer dataset has 6 flow configurations, compared to 90 in the training dataset. 
Therefore, the former has $15\times$ fewer flow configurations than the latter, dramatically reducing the overall data collection at fine-scale discretizations.

\emph{\textbf{Validation dataset}}. We collect validation datasets to control the under- and over-fitting of the network during both the pre-training and fine-tuning phases. 
The validation dataset consists of flow around three different unseen-in-training geometries: an ellipse $AR=0.22$, a symmetric NACA0012 airfoil, and a cylinder. 
For each geometry, we consider 3 flow variations -- two $\alpha$ angles and one $\theta$ angle, chosen randomly as in the training dataset for a total of nine flow configurations. 
Note that we collect the same validation dataset at all resolutions, from $64\times256$ to $2048\times 2048$.

\begin{figure}[htbp]
\centering
\small
\includegraphics[width=0.92\columnwidth]{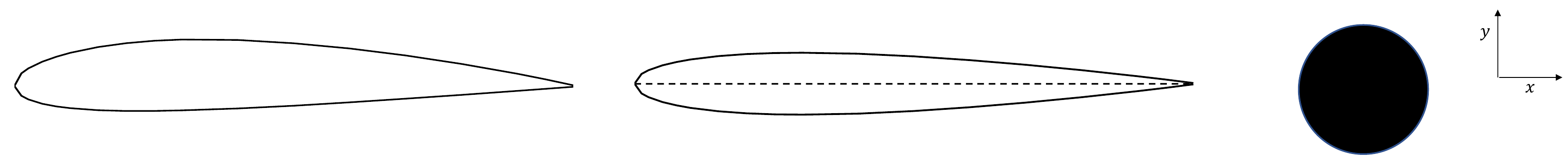}
	\caption{\small Non-symmetric NACA1412 airfoil (left), symmetric NACA0015 airfoil (center), and cylinder (right) as test geometries. The last two digits in the 4-digit NACA denomination represent the maximum thickness percentage of the chamber of the airfoil with respect to the airfoil's chord (dashed line).}
\label{fig:test_airfoils}
\end{figure}

\emph{\textbf{Test dataset}}. We collect a new dataset to evaluate \name's performance in accelerating high-resolution turbulent flow simulations. 
The test datasets consist of flow around four geometries: a non-symmetric NACA1412 airfoil, a symmetric NACA0015 airfoil, an ellipse $AR=0.3$, and a cylinder. The airfoil and cylinder geometries are shown in Figure~\ref{fig:test_airfoils}. 
The test dataset contains geometries \emph{unseen} during the pre-training or transfer learning phases. 
Nonetheless, some geometries are, a priori, more challenging than others. 
For example, the non-symmetric NACA1412 airfoil has three unique features distinct from the training dataset: the flat trailing edge, the non-symmetry, and the chamber thickness (12\% of the chord of the airfoil). 
Comparing the airfoil to the ellipse in the test set, the only feature unseen during the training phase is its chamber thickness ($AR=0.3$). 
Moreover, even though a cylinder is a special case of an ellipse from a geometrical perspective, the physics in the rear of the cylinder has a large recirculation area not present in the training flows.
For each geometry, we consider 2 flow variants: one $\alpha$ angle and one $\theta$ angle. Table~\ref{tab:results} summarizes the different test cases. The test dataset is collected at all resolutions. 

\vspace{-1.5em}
\subsection{Physics Solver}
\label{sec:openfoam}
The training, transfer, validation, and test datasets are generated by solving the RANS equations with the Spalart-Allmaras one-equation model \cite{SpalartAllmaras}. 
We use the incompressible solver \texttt{simpleFoam} from OpenFOAM v8 as the \emph{physics solver} that implements the semi-implicit method for pressure-linked equations (SIMPLE) algorithm.
A residual value below \num{1e-4} is the criteria to consider the training simulations converged. 
We adhere to tolerances that are extended practice in CFD simulations~\cite{ling}. 
We use the \texttt{GAMG} solver for computing the pressure at every iteration, with a tolerance of \num{1e-8}.
The \texttt{smoothSolver} and the \texttt{GaussSeidel} smoother compute both the velocity and modified eddy viscosity.

\textbf{\emph{Architecture and Libraries.}} All the OpenFOAM simulations are run in parallel on a dual-socket Intel Xeon Gold 6148 using double precision.
Each socket has 20 cores, for a total of 40 cores.
We use the OpenMPI implementation of MPI integrated with OpenFOAM v8 that is optimized for shared-memory communication.
The grid domain is decomposed into 40 partitions using the integrated Scotch partitioner and each partition is assigned to 1 MPI process that is pinned to a single core. 
We set the \texttt{numactl -localalloc} flag to bind each MPI process to its local memory.

\subsection{Coarse Model Training}
\label{sec:training}
We train the CNN in Figure~\ref{fig:network}\footnote{We also considered a 10- and 12-layer CNN. However, adding additional layers did not improve performance.} using the training dataset described in Section~\ref{sec:datasets} using double precision.
We implement the CNN using Keras~\cite{keras} and perform distributed training on four Tesla V100 GPUs connected with PCIe, using the TensorFlow 2.4 backend. 
No specific initialization is used in training. 
The batch size is 64, the optimizer is Adam, and the loss function is mean squared error (MSE). 
The learning rate is set to \num{1e-4} with no decay for all training. 
The training is stopped using the EarlyStopping Keras callback~\cite{EarlyStopping} by monitoring the validation loss with patience of 6 epochs. 
After 41 epochs, the training loss reaches \num{7e-4} and validation loss
reaches a value of \num{8e-4}.

%% file: tables/TableMeshDataset.tex
\newcommand*\TableMeshDataset{
	\begin{table}[htbp]
		\small
	\begin{center}
\renewcommand{\arraystretch}{0.95}
		\begin{tabular}{ll|c|c}

\toprule

			Datasets&  & Low resolution & Higher resolutions 

			\\[0.7ex]
\hline
			&&&\\[-1em]

			\multirow{3}{*}{ Training } & NoG &   10      &  \multirow{3}{*}{\tikzcirclered{1.8pt}}\\
			    & NoFC  &  90   &                     \\     
			    & Type &  ellipses &                \\            [0.7ex]
\hline
			&&&\\[-1em]
			\multirow{3}{*}{ Transfer } & NoG &    \multirow{3}{*}{\tikzcirclered{1.8pt}}      &  1\\
			    & NoFC  &     &     6\\     
			    & Type &     &       ellipse $AR=0.1$\\            [0.7ex]
			    \hline

			\multirow{3}{*}{ Validation } & NoG &    \multicolumn{2}{c}{3}\\
			    & NoFC  &   \multicolumn{2}{c}{9}\\     
			    & Type  &    \multicolumn{2}{c}{ \makecell[c]{NACA0012, ellipse $AR=0.22$, cylinder}}\\        [0.7ex]
\hline
			\multirow{3}{*}{ Test } & NoG &    \multicolumn{2}{c}{4}\\
			    & NoFC  &   \multicolumn{2}{c}{8}\\    
			    & Type  &    \multicolumn{2}{c}{ \makecell[c]{NACA1412, NACA0015, \\ ellipse $AR=0.3$, cylinder}}\\            
\bottomrule
\end{tabular}
		\caption{\small Summary of datasets. The training dataset is from low-resolution simulations. At all high resolutions we collect identical transfer, validation, and test datasets. NoG is for the number of different geometries, and NoFC is for the number of flow configurations (\ie total number of simulations). \label{tab:datasets}}
\end{center}
\end{table}

}

%% file: text/results.tex
\vspace{-0.5em}
\section{Results and Discussion}
\label{sec:results}
After pre-training and validating the CM, we perform fine-tuning using both transfer learning approaches (\ie OSTL and ITL) and evaluate \name's ability for super-resolution at various target high-resolution discretizations. 
We start by empirically demonstrating the inefficiency of CM without fine-tuning to recover high-resolution turbulent flows, especially at fine-scale discretizations. 
Then, we evaluate \name's TL and its ability to generalize to geometries unseen during the pre-training and fine-tuning phases. 
Finally, we evaluate its performance in reconstructing high-resolution turbulent flows with respect to the OpenFOAM physics solver. 
Besides, we also compare \name\ against the baseline model (aka oracle), which performs full training with a large training dataset collected at higher resolutions.

\input{tables/TableResults}

\input{tables/TableWarmup}

\subsection{Validation Loss}
\label{subsec:results:loss}
\begin{figure}[htbp]
\centering
\small
\includegraphics[width=0.95\columnwidth]{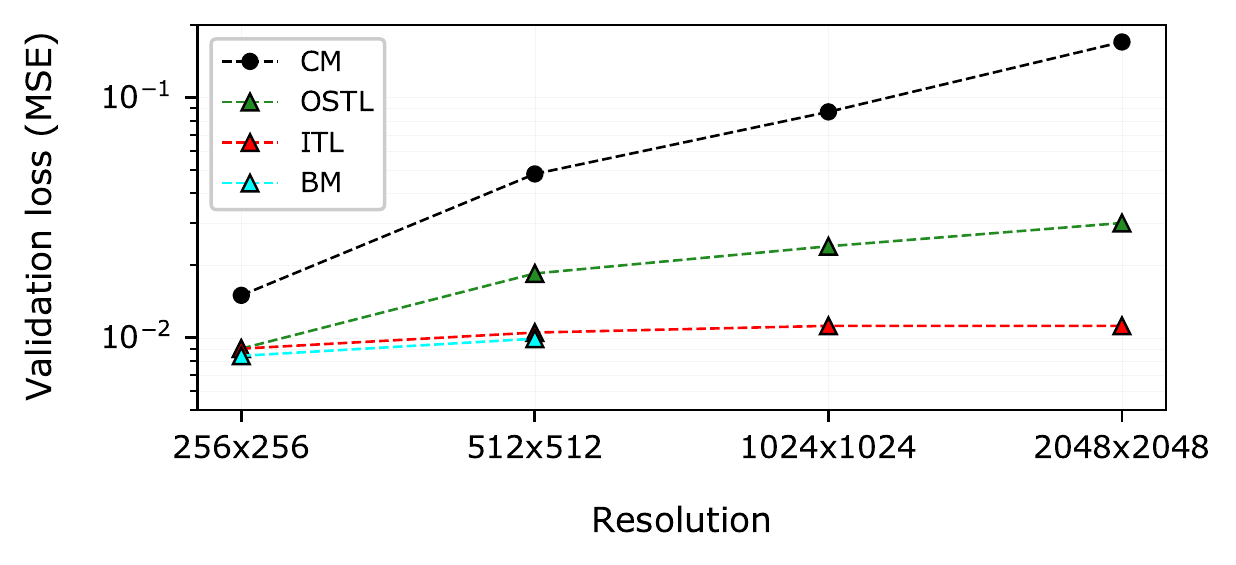}
	\vspace{-2em}
	\caption{\small Coarse model (CM), one-shot transfer learning (OSTL), incremental transfer learning (ITL), and baseline model (BM) losses on the validation dataset at every target resolution.}

\label{fig:losses}
\end{figure}

One of our objectives is to maintain prediction accuracy across discretizations to build a resolution-invariant DL algorithm. 
Figure~\ref{fig:losses} shows the validation loss of the different models with increasing resolution size.  

\textbf{\emph{CM loss.}} 
We observe that the validation loss of CM increases significantly with increasing resolution size from \num{1.5e-3} at $256\times256$ to \num{2.1e-1} at $2048\times2048$. 
These results indicate that CM trained with low-resolution data is unable to recover high-resolution flow fields. 
This lack of fidelity is because coarse discretizations learned with CM are incapable of capturing sharp gradients and resolving flow instabilities prevalent at fine discretizations. 
Hence, CM alone is inadequate for super-resolution. 

\textbf{\emph{OSTL loss.}} To augment the prediction capabilities of CM, we perform one-shot TL from CM to the target high-resolution using the transfer dataset as
discussed in Section~\ref{sec:methodology}. 
We set the learning rate to one order of magnitude lower than the training learning rate for CM (from \num{1e-4} to \num{1e-5}), and train for \emph{only} one or two epochs. 
The loss of the fine-tuned OSTL model on the validation dataset significantly reduces at all target resolutions compared to CM as seen from Figure~\ref{fig:losses}. 
At $256\times 256$, OSTL reduces the validation loss from \num{1.5e-2} for CM to \num{9e-3}, resulting in 60\% overall reduction.  
We observe a similar trend with subtle improvements at larger resolutions.
At $2048\times 2048$ the validation loss reduces from \num{2e-1} for CM to \num{3e-2} for OSTL -- almost one order of magnitude. 

We make two main observations from the OSTL results. 
First, the gap in the validation losses between OSTL and CM increases with the grid size. 
Coarse discretizations do not contribute sufficient points in the domain to obtain accurate solutions. 
In areas of strong gradients, 
fine discretizations shape the flow field very differently from coarse discretizations as seen from Figures~\ref{fig:results_NACA0015} and \ref{fig:results_NACA1412} for the NACA airfoils.
The finer the discretization, the more distinct the resulting flow field is from the baseline low resolutions. Therefore, when we transfer from CM
to very high resolutions (\eg $2048\times 2048$), the new flow field seen during the transfer phase produces sizable changes to the weights of the network compared to transferring to "mid-resolutions" (\eg $256 \times 256$). 
Second, this divergence of the flow features
between low- and high-resolution discretizations also explains why the OSTL validation loss increases with the grid size. 
Although OSTL substantially improves the accuracy of super-resolution, it still does not achieve resolution-invariance. 

To alleviate these limitations, we consider three alternatives. 
First, to train for more epochs during fine-tuning. 
However, this comes with the risk of overfitting to the unique geometry in the transfer dataset, which is undesirable. 
Second, to include more geometries in the transfer dataset. 
This approach is also detrimental because it would require substantial data collection at high resolutions. 
Third, use incremental transfer learning (ITL).
This approach is promising as the model is further fine-tuned with data at intermediate resolutions until the target discretization. 
We choose a step size of 1 to avoid the drawbacks of OSTL described above and improve the accuracy of the network at very high resolutions while still reducing the overall data collection.

\textbf{\emph{ITL loss.}} \name\ does incremental transfer learning using the same approach as in OSTL.
Figure~\ref{fig:losses} plots the error of \name\ after ITL on the validation dataset at each target resolution. 
\name's OSTL and ITL approaches produce different validation loss values at every target resolution except $256\times256$. ITL improves the generalizability of the network.
The validation loss drops by half at $512\times512$ and $3\times$ at $2048\times2048$ compared to OSTL. 
ITL reaches a loss that is invariant to the resolution: the validation loss remains constant at around \num{1e-2} for every target discretization. 
Because the transfer dataset at each resolution is $15\times$ smaller than the training dataset, we still learn incrementally using far less data. 
Most importantly, ITL achieves similar accuracy to the baseline model or oracle trained using a dataset as large as the pre-training dataset for CM at $256\times256$ and $512\times512$.
This result is particularly notable because ITL reaches oracle-level accuracy without the need for exhaustive training with large input sizes and large-scale high-resolution datasets.
We present a more detailed performance comparison against the oracle in Section~\ref{subsec:results:baseline}.

\subsection{Performance Analysis}
\label{subsec:results:performance}
We now study \name's performance in super-resolution of turbulent flows. 
Recall that \name's pre-training consists of training the base network with a large number of inputs of low-resolution data. 
The transfer phase adds a minimum amount of high-resolution data for fine-tuning the network. 
Therefore, in addition to performance, we also evaluate the ability of \name\ in recovering the \emph{same} high-resolution turbulent flow solution as the physics solver.

    \begin{table*}[htbp]
	\centering
\scriptsize
\colorbox{lightgray}{
\setlength{\tabcolsep}{2.8pt}
\renewcommand{\arraystretch}{0.9}
	\begin{tabular}{l|l|cc|cccc|ccccc|cccc|cccc}

		    \multicolumn{2}{c}{}
		    & \multicolumn{2}{c}{$64\times 256$}
		    & \multicolumn{4}{c}{$256\times256$}
		    & \multicolumn{5}{c}{$512\times512$}
		    & \multicolumn{4}{c}{$1024\times1024$}
		    & \multicolumn{4}{c}{$2048\times2048$}\\[0.7ex]
		    \cline{1-21}
		    \multicolumn{2}{c}{}&&&&&&&&&&&&&&&&&&&\\[-0.7em]

		    \multicolumn{2}{l}{Test case}
		    & PS
		    & C-SN
		    & PS
		    & C-SN
		    & O-SN
		    & BM
		    & PS
		    & C-SN
		    & O-SN
		    & I-SN
		    & BM
		    & PS
		    & C-SN
		    & O-SN
		    & I-SN
		    & PS
		    & C-SN
		    & O-SN
		    & I-SN\\[0.7ex]

\hline
		    &&&&&&&&&&&&&&&&&&&&\\[-0.7em]
		    
		    \input{tables/NACA1412theta}
		    \hline
		    &&&&&&&&&&&&&&&&&&&&\\[-0.7em]
		    \input{tables/NACA1412alpha}
		    \hline
		    &&&&&&&&&&&&&&&&&&&&\\[-0.7em]
       		    \input{tables/naca0015theta}
		    \hline
		    &&&&&&&&&&&&&&&&&&&&\\[-0.7em]
       		    \input{tables/naca0015alpha}
		    \hline
		    &&&&&&&&&&&&&&&&&&&&\\[-0.7em]
		    \input{tables/ellipsetheta}

                    \hline
		    &&&&&&&&&&&&&&&&&&&&\\[-0.7em]
		    \input{tables/ellipsealpha}
		    \hline
		    &&&&&&&&&&&&&&&&&&&&\\[-0.7em]
		    \input{tables/cylindertheta}
		    \hline
		    &&&&&&&&&&&&&&&&&&&&\\[-0.7em]
		    \input{tables/cylinderalpha}
		    \bottomrule
  \end{tabular}}
		\caption{\small Summary of the performance results. TTC is the time-to-convergence and ITC is the number of iterations-to-convergence of the physics solver (PS), C-SURFNet (C-SN), O-SURFNet (O-SN), I-SURFNet (I-SN) and the Baseline Model (BM). The speedup of all \name\ models is calculated with respect to the physics solver. \label{tab:results}}
\end{table*}

\TableWarmup

We test \name\ in simulations of turbulent flow around 4 unseen-in-training geometries at 2 flow configurations each, for a total of 8 test cases at all target resolutions. 
Table~\ref{tab:results} presents the comparison against the OpenFOAM physics solver. 
\name\ creates three models -- (i) low-resolution coarse model, (ii) OSTL model, and (iii) ITL model. 
Therefore, at every target resolution, we evaluate the time-to-convergence (TTC) using each model, namely: C-SURFNet, O-SURFNet, and I-SURFNet. 
We compare the TTC of each one of these models with the TTC of the physics solver (PS) and baseline model (BM)\footnote{A comparison is presented against the BM (or oracle) for only $256\times256$ and $512\times512$ due to the computational cost of collecting large datasets and training at higher resolutions.}. 

The TTC of the PS is computed by using in tandem two popular convergence criteria in the CFD literature~\cite{ling}: (1) the residual of each flow variable drops 4 to 6 orders of magnitude and (2) by monitoring when physical quantities reach steady-state. 
In contrast, SURFNet reaches convergence in three stages. 
First there is \emph{warmup} (W), where we start the simulation with the PS and let the residual drop between one and two orders of magnitude for each variable. This is sufficient for the fluid parameters close to the physical boundaries to capture the geometry of the new problem. 
Next there is \emph{inference} (I), where we use these intermediate flow variables as input to the CNN, which infers the steady-state.
Finally, the CNN's output is constrained with the PS in \emph{refinement} to reach the same convergence criteria as the PS~\cite{cfdnet}. 
The TTC of \name\ is the sum of the time spent in the three stages.
The three stages are run in parallel on the CPU described in
Section~\ref{sec:openfoam} for a fair comparison against OpenFOAM's solver (which doesn't support GPU acceleration). Table~\ref{tab:warmup} presents the time spent at W and I and the number of iterations in W. The refinement time can be found by adding W and I and substracting this sum from the TTC of Table~\ref{tab:results}.

\emph{\textbf{C-SURFNet vs. physics solver}}. The first column in Table~\ref{tab:results} presents the eight cases in our test dataset. 
The results of C-SURFNet at $64\times256$ indicate good generalization capacity at the lowest resolution to \emph{unseen} geometries. 
We observe consistent speedups around $2-2.1\times$, independent of the geometry or flow configuration. 
On the other hand, the first row shows the results for the first test case, flow around a NACA1412 at $\theta=5^{\circ}$ at different spatial resolutions. 
Although we observe significant speedups at low resolutions and C-SURFNet exhibits improved TTC compared to the PS, its performance gain also degrades consistently with increasing discretizations.
The speedup drops from $2.1\times$ at $64\times256$ to $1\times$ at $2048\times2048$, where it's no better than the PS. 
We observe a similar trend across all the test cases. 
These results are in accordance with the observations in Section~\ref{subsec:results:loss}, where the CM that has learned from only low-resolution inputs is incapable of predicting accurate high-resolution solutions. 

\begin{figure}[htbp]
	\centering
	\includegraphics[width=0.95\columnwidth]{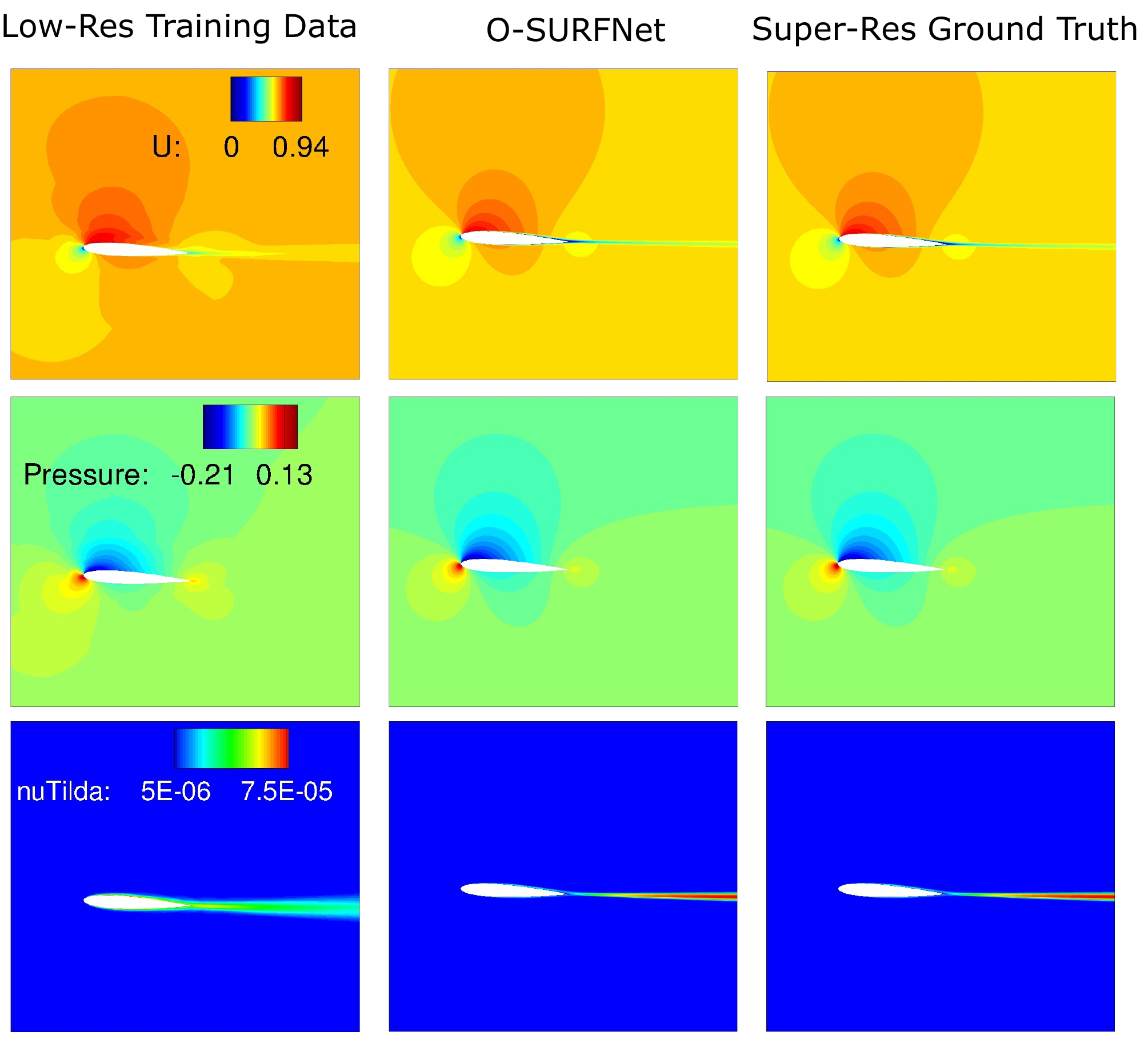}	
	\caption{\small Velocity in \si{\meter\per\second} (top), kinematic pressure in \si{\meter^2\per\second^2} (middle), and modified eddy viscosity in \si{\meter^2\per\second} (bottom) around the NACA 0015 airfoil at $\theta=3^{\circ}$, \Rey\ = \num{6e5}. Comparison between the low-resolution ($64\times256$) training data to train the CM (left); O-SURFNet's output after the refinement phase at $2048\times 2048$ (middle), and the ground truth OpenFOAM's solution at $2048\times 2048$ (right). 
	\vspace{-1em}
	\label{fig:results_NACA0015}}
\end{figure}

\emph{\textbf{O-SURFNet vs. physics solver}}. In Table~\ref{tab:results}, all high resolutions have an O-SURFNet (O-SN) column, and we make three main observations from the results. 
First, O-SURFNet outperforms C-SURFNet for all test cases and resolutions. 
For instance, for the symmetric NACA0015 at $\theta=3^{\circ}$ at $2048\times2048$, the TTC of the PS is 88.7 minutes. 
O-SURFNet reduces this time to 55.4 minutes resulting in a $1.6\times$ speedup (C-SURFNet achieved no performance gain). 
Figure~\ref{fig:results_NACA0015} shows the corresponding qualitative results of O-SURFNet in the task of super-resolution. 
It resolves the fields of all fluid variables - velocity, pressure, and the modified eddy viscosity - at $2048\times2048$ faster than the PS while being pre-trained with low-resolution data with \emph{only} fine-tuning at the highest spatial resolution. 
Second, O-SURFNet's performance is better at lower resolutions than higher resolutions. 
For the same NACA0015 test case, at $1024\times1024$, $512\times512$, and $256\times256$, the performance gains are $1.7\times$, $1.8\times$, and $2.1\times$ respectively. 
We observe a similar trend of decaying performance with increasing resolutions across all test cases similar to the coarse model, albeit not to the same extent. 
This is in line with the results presented in Section~\ref{subsec:results:loss} where OSTL losses increase the more dissimilar the target flow field is from the low-resolution flow. 
OSTL at $256\times256$ discretization achieves a $2\times$ speedup irrespective of the test case, demonstrating its potential to generalize to higher resolutions provided the target flow features have sufficient overlap with the tiny resolution used for pre-training the CM from which it fine-tunes. 
However, this is also an indication that OSTL is incapable of achieving resolution-invariance. 
Third, the speedups achieved by O-SURFNet remain constant and stable among test cases (for instance, we observe speedups around $1.8\times$ at $512 \times 512$), indicating that fine-tuning the model did not result in overfitting to the transfer geometry. 
Neither CM nor OSTL yield overfitted models and exhibit stable generalization capacities.

\begin{figure}[htbp]
	\centering
  	\includegraphics[width=0.95\columnwidth]{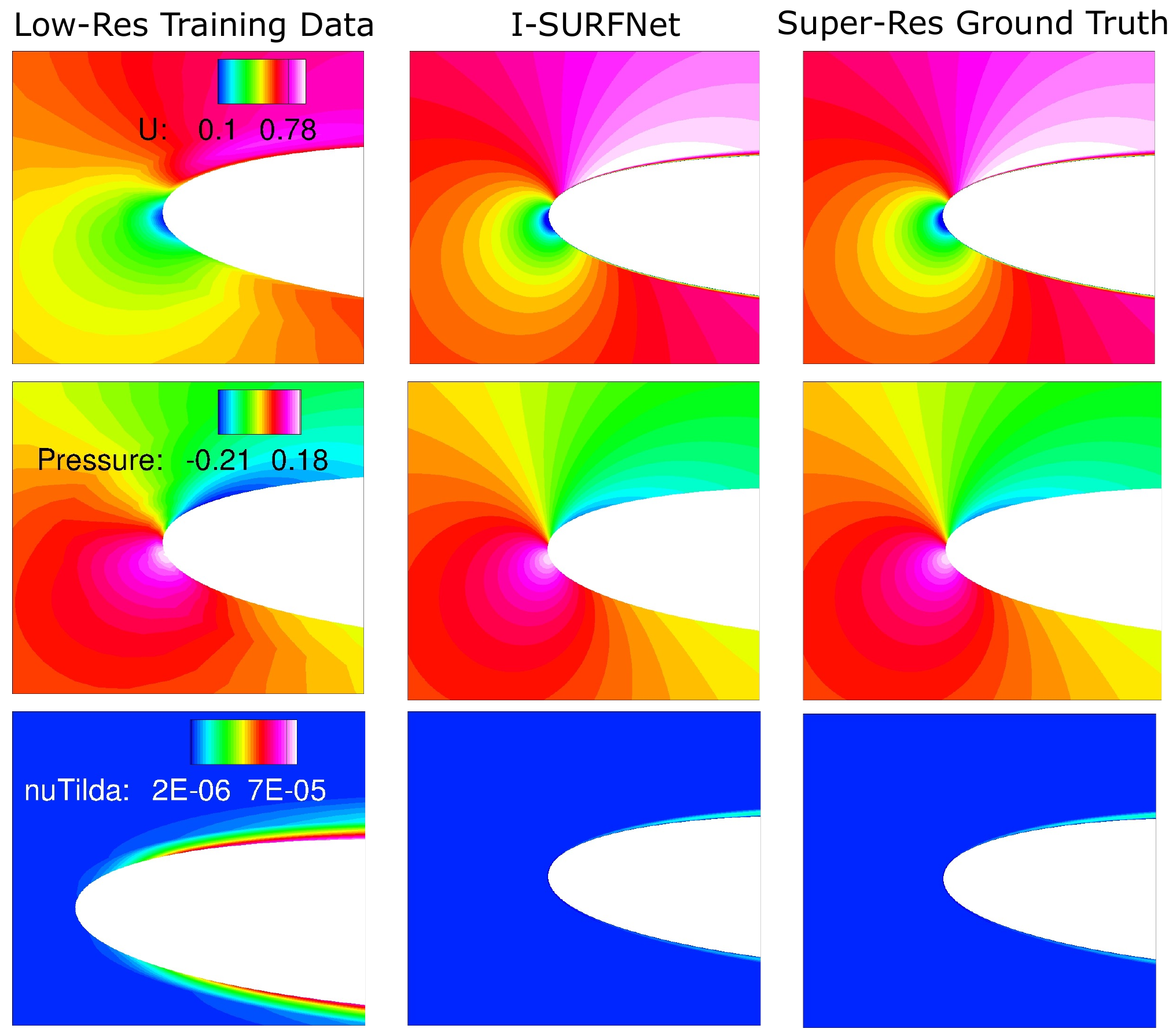}
	\caption{\small Detail of the velocity in \si{\meter\per\second} (top), kinematic pressure in \si{\meter^2\per\second^2} (middle), and modified eddy viscosity in \si{\meter^2\per\second} (bottom) at the nose of the nonsymmetric NACA 1412 airfoil at $\theta=5^{\circ}$, \Rey\ = \num{6e5}. Comparison between the low-resolution ($64\times256$) training data to train the coarse model (left), I-SURFNet's output after the refinement phase at $2048\times 2048$ (middle), and the ground truth OpenFOAM's solution at $2048\times 2048$ (right).
	\vspace{-1em}
\label{fig:results_NACA1412}}	
\end{figure}

\emph{\textbf{I-SURFNet vs. physics solver}}. O-SURFNet's accuracy decreases as the resolution increases. 
An ideal solution is one that maintains accuracy (\ie resolution-invariant) while simultaneously requiring limited computational resources. 
In Table~\ref{tab:results}, we observe that across the board (\ie all test geometries and flow configurations unseen during pre-training and TL), I-SURFNet achieves a $2 - 2.1\times$ speedup against the PS. 
Not only does I-SURFNet maintain the performance gain over PS across the different test cases but also across all target resolutions demonstrating both \emph{generalization} and \emph{resolution-invariance}. 
Since ITL incrementally fine-tunes the model, it requires more data than OSTL.
However, fine-tuning is done on the same geometry across resolutions with $15\times$ lesser data, thereby significantly reducing the overall data collection at high spatial discretizations (compared to prior approaches that require considerably more high-resolution data).
Figure~\ref{fig:results_NACA1412} shows the qualitative results of I-SURFNet's flow solution around the nose of the non-symmetric NACA1412 airfoil at $\theta=5^{\circ}$.
I-SURFNet successfully recovers high-resolution turbulent flow simulations on a geometry with at least three distinct features (\ie flat trailing edge, non-symmetry, and different chamber thickness) not present in the training or transfer datasets. 
This validates that \name\ pre-trained with low-resolution data with only fine-tuning can generalize to unseen geometries.
The largest target resolution studied (\ie $2048\times2048$) is \textbf{256$\times$} the size of the tiny discretization (\ie $64\times256$) used in pre-training the CM to stress SURFNet's ability in super-resolution. It is guaranteed to converge to a unique solution as long as the problem is well-posed~\cite{chaosinsystems}. 

We additionally explored the effect of the required size of the spatial step size to maintain accuracy for ITL. 
By incrementally fine-tuning with a step size of 1 up to $2048\times2048$, the performance gain is consistently $2\times$. 
This gain is the best possible achievable as observed by the results of the oracle or BM that is fully trained at $256\times256$ and $512\times512$ resolutions.
By incrementally transferring with a step size of 2 up to $2048\times2048$, the performance gain drops to $1.7-1.8\times$.
So, we conclude that to achieve oracle-level accuracy and performance, the ideal step size for ITL is 1.

\subsection{SURFNet vs. Oracle}
\label{subsec:results:baseline}
We now compare \name\ with the oracle or baseline model (BM) at $256\times256$ and $512\times512$ target discretizations. 
We define BM as conducting full-scale data collection \emph{and} training at the higher resolutions. 
Specifically, a dataset as large as the training dataset outlined in Table~\ref{tab:datasets} collected at the tiny resolution ($64\times 256$) for pre-training the CM is collected for training the CNN at the two target discretizations. 
The CNN is trained using a learning rate of \num{1e-4} and a batch size of $32$ and $16$, respectively. 
Figure~\ref{fig:losses} and Table~\ref{tab:results} show that the loss on the validation dataset and speedup compared to the OpenFOAM physics solver achieved by ITL is similar to that of the BM.
To understand how much time \name\ saves with respect to the oracle, Table~\ref{tab:baseline} compares the data collection and the training time of \name\ with that of BM. 
\name's data collection time is the sum of the time spent collecting low-resolution training data and the transfer datasets at higher resolutions.
Similarly, the total training time includes pre-training the CM and the time spent fine-tuning at the target higher resolution.

In terms of data collection time, \name\ takes 0.4 hrs (0.33 + 0.06) vs. 1 hr for BM at $256\times256$. 
At $512\times 512$, \name's data collection time is 0.8 hrs (0.33 + 0.06 + 0.41) vs. 3.86 hrs for BM. 
Training at $256\times 256$ and $512\times 512$ takes 2.56 hrs and 3 hrs for SURFNet vs. 9.25 hrs and 38 hrs for BM, respectively. 
Overall, SURFNet reduces the combined data collection and training time by $3.6\times$ and $10.2\times$, respectively, while achieving similar performance gain and accuracy as BM. 
Note that the computational advantage of \name\ increases with increasing resolution size.
Moreover, these results highlight the impracticability of performing exhaustive data collection and training at $1024 \times 1024$ and beyond, underscoring the impact and potential of TL (specifically ITL) in enabling super-resolution of complex turbulent flows. 

\input{tables/TableBaseline}

\TableBaseline

%% file: tables/TableResults.tex
\newcommand*\TableResults{
    \begin{table*}[htbp]
	\centering
\scriptsize
\colorbox{lightgray}{
\setlength{\tabcolsep}{2.8pt}
\renewcommand{\arraystretch}{0.9}
	\begin{tabular}{l|l|cc|cccc|ccccc|cccc|cccc}

		    \multicolumn{2}{c}{}
		    & \multicolumn{2}{c}{$64\times 256$}
		    & \multicolumn{4}{c}{$256\times256$}
		    & \multicolumn{5}{c}{$512\times512$}
		    & \multicolumn{4}{c}{$1024\times1024$}
		    & \multicolumn{4}{c}{$2048\times2048$}\\[0.7ex]
		    \cline{1-21}
		    \multicolumn{2}{c}{}&&&&&&&&&&&&&&&&&&&\\[-0.7em]

		    \multicolumn{2}{l}{Test case}
		    & PS
		    & C-SN
		    & PS
		    & C-SN
		    & O-SN
		    & BM
		    & PS
		    & C-SN
		    & O-SN
		    & I-SN
		    & BM
		    & PS
		    & C-SN
		    & O-SN
		    & I-SN
		    & PS
		    & C-SN
		    & O-SN
		    & I-SN\\[0.7ex]

\hline
		    &&&&&&&&&&&&&&&&&&&&\\[-0.7em]
		    
		    \input{tables/NACA1412theta}
		    \hline
		    &&&&&&&&&&&&&&&&&&&&\\[-0.7em]
		    \input{tables/NACA1412alpha}
		    \hline
		    &&&&&&&&&&&&&&&&&&&&\\[-0.7em]
       		    \input{tables/naca0015theta}
		    \hline
		    &&&&&&&&&&&&&&&&&&&&\\[-0.7em]
       		    \input{tables/naca0015alpha}
		    \hline
		    &&&&&&&&&&&&&&&&&&&&\\[-0.7em]
		    \input{tables/ellipsetheta}

                    \hline
		    &&&&&&&&&&&&&&&&&&&&\\[-0.7em]
		    \input{tables/ellipsealpha}
		    \hline
		    &&&&&&&&&&&&&&&&&&&&\\[-0.7em]
		    \input{tables/cylindertheta}
		    \hline
		    &&&&&&&&&&&&&&&&&&&&\\[-0.7em]
		    \input{tables/cylinderalpha}
		    \bottomrule
  \end{tabular}}
		\caption{\small Summary of the performance results. TTC is the time-to-convergence and ITC is the number of iterations-to-convergence of the physics solver (PS), C-SURFNet (C-SN), O-SURFNet (O-SN), I-SURFNet (I-SN) and the Baseline Model (BM). The speedup of all \name\ models is calculated with respect to the physics solver. \label{tab:results}}
\end{table*}
}

%% file: tables/NACA1412theta.tex
\multirow{3}{*}{ \makecell[l]{NACA 1412\\$\theta=5^{\circ}$}}
& TTC (min)
&0.16&	0.08&	0.4&	0.22&	0.2&  0.2  & 	2.4&	1.7&	1.3&	1.2&	1.2& 17.2&	14.3&	10.1&	8.6&	95&	95&	59.4&	47.5

\\[0.7ex] 
 & ITC

&1865&	825&	3636	&1874&	1672& 1669 &7273&	5018&	3864&	3460& 3473	&10118&	8259&	5779&	4887	&12338&	12183	&7556	&6014

\\[0.7ex] 
& Speedup
& 1$\times$
& \best{2.1$\times$}  

& 1$\times$
& \worst{1.8$\times$}  
& \best{2$\times$}    
& \best{2$\times$}    

& 1$\times$
& \worst{1.4$\times$}  
& 1.8$\times$         
& \best{2$\times$}   
& \best{2$\times$}   

& 1$\times$
& \worst{1.2$\times$}   
& 1.7$\times$   
& \best{2$\times$}   

& 1$\times$
& \worst{1$\times$}   
& 1.6$\times$   
& \best{2$\times$}   
\\[0.7ex] 

%% file: tables/NACA1412alpha.tex
\multirow{3}{*}{ \makecell[l]{NACA 1412\\$\alpha=6^{\circ}$}}
& TTC (min)

&0.19&	0.09	&0.53	&0.29	&0.27 & 0.25 	&3.3	&2.5	&1.9	&1.7 &1.7	&19.3	&16.1	&12.06	&9.7	&98.5	&89.5	&61.6	&49.3
\\[0.7ex] 

& ITC

&2215&	991&	4818&	2531&	2263&   2199 &10000&	7516&	5706&	4823& 4830  &11353	&9289&	6506&	5504&	12792&	11474&	7840&	6241

\\[0.7ex] 

& Speedup 
& 1$\times$
& \best{2.1$\times$}

& 1$\times$
& \worst{1.8$\times$}
& \best{2$\times$}
& \best{2.1$\times$}

& 1$\times$
& \worst{1.3$\times$}
& 1.7$\times$
& \best{2$\times$}
& \best{2$\times$}

& 1$\times$
& \worst{1.2$\times$}
& 1.6$\times$
& \best{2$\times$}

& 1$\times$
& \worst{1.1$\times$}
& 1.5$\times$
& \best{2$\times$}

\\[0.7ex] 

%% file: tables/naca0015theta.tex
\multirow{3}{*}{ \makecell[l]{NACA 0015\\$\theta=3^{\circ}$}}
& TTC (min)

&0.15&	0.07&	0.34&	0.18&	0.16& 0.17	&2.5&	1.7	&1.4	&1.2& 1.2&	15.8&	12.2&	8.8&	7.9&	88.7&	88.7	&55.4	&44.4

\\[0.7ex] 
& ITC

& 1748&	769	&3091&	1481	&1326& 1349	&7576&	4874&	4032	&3431 & 3424	&9294	&6977	&4991&	4475	&11519	&11365&	7045&	5605
\\[0.7ex]

& Speedup 
& 1$\times$
& \best{2.1$\times$}

& 1$\times$
& \worst{1.9$\times$}
& \best{2.1$\times$}
& \best{2$\times$}

& 1$\times$
& \worst{1.5$\times$}
& 1.8$\times$
& \best{2.1$\times$}
& \best{2.1$\times$}

& 1$\times$
& \worst{1.3$\times$}
& 1.7$\times$
& \best{2$\times$}

& 1$\times$
& \worst{1$\times$}
& 1.6$\times$
& \best{2$\times$}\\[0.7ex]

%% file: tables/naca0015alpha.tex
\multirow{3}{*}{ \makecell[l]{NACA 0015\\$\theta=\alpha=0^{\circ}$\\}}
& TTC (min)

& 0.13&	0.06&	0.32&	0.18&	0.16& 0.16 &	2.3&	1.6&	1.3&	1.2& 1.2 &	14.5&	10.4&	8.5&	7.3&	82.4&	82.4&	51.5&	41.2   

\\[0.7ex] 
 & ITC

&1515&	626&	2909&	1470&	1308&1311	&6970&	4802&	3696&	3308&3300&	8529&	5920&	4845&	4093&	10701&	10546&	6533&	5196

\\[0.7ex] 

&Speedup
& 1$\times$
& \best{2.2$\times$}  

& 1$\times$
& \worst{1.8$\times$}  
& \best{2$\times$}   
& \best{2$\times$}   

& 1$\times$
& \worst{1.4$\times$}  
& 1.8$\times$         
& \best{2$\times$}   
& \best{2$\times$}   

& 1$\times$
& \worst{1.2$\times$}  
& 1.7$\times$   
& \best{2$\times$}   

& 1$\times$
& \worst{1$\times$}   
& 1.6$\times$   
& \best{2$\times$}   

\\[0.7ex]

%% file: tables/ellipsetheta.tex
\multirow{3}{*}{ \makecell[l]{Ellipse \\$AR=0.3$ \\$\theta=5^{\circ}$}}
& TTC (min)

& 0.22&	0.1&	0.61&	0.36	&0.31	&0.3&4.1&	3.2&	2.4&	2.1&2.1&	20.8&	17.3&	13.0&	10.4&	107.2&	97.5&	67.0&	53.6

\\[0.7ex] 
 & ITC

& 2564&	1158	&5545&	3116	&2627& 2615 &	12424&	9381&	7132&	6036& 6039 &	12235	&10024&	7475&	5946&	13922&	12502&	8546&	6806

\\[0.7ex] 
& Speedup
& 1$\times$
& \best{2.1$\times$}

& 1$\times$
& \worst{1.8$\times$}  
& \best{2$\times$}    
& \best{2$\times$}    

& 1$\times$
& \worst{1.3$\times$}  
& 1.7$\times$         
& \best{2$\times$}   
& \best{2$\times$}   

& 1$\times$
& \worst{1.2$\times$}   
& 1.6$\times$   
& \best{2$\times$}   

& 1$\times$
& \worst{1$\times$}  
& 1.5$\times$   
& \best{2$\times$}   
\\[0.7ex]

%% file: tables/ellipsealpha.tex
\multirow{3}{*}{ \makecell[l]{Ellipse\\$AR=0.3$\\$\alpha=1^{\circ}$}}
& TTC (min)

& 0.25	& 0.13	& 0.64	& 0.36	& 0.32	& 0.33 & 4.6& 	3.3& 	2.6& 	2.3& 2.3 & 22.3& 	17.2	& 13.1& 	11.2& 	114.7& 	104.3& 	71.7& 	57.4

\\[0.7ex] 
 & ITC

&2914&	1394	&5818&	3086&	2763& 2777 &	13939&	9780&	7568&	6793& 6790 &	13118&	9918&	7544&	6387&	14896&	13387&	9155&	7293

\\[0.7ex] 
& Speedup

& 1$\times$
& \best{2$\times$}  

& 1$\times$
& \worst{1.8$\times$}  
& \best{2$\times$}    
& \best{2$\times$}    

& 1$\times$
& \worst{1.4$\times$} 
& 1.8$\times$        
& \best{2$\times$}   
& \best{2$\times$}   

& 1$\times$
& \worst{1.3$\times$}  
& 1.7$\times$   
& \best{2$\times$}   

& 1$\times$
& \worst{1.1$\times$}  
& 1.6$\times$   
& \best{2$\times$}   
\\[0.7ex]

%% file: tables/cylindertheta.tex
\multirow{3}{*}{ \makecell[l]{Cylinder\\$\theta=0^{\circ}$}}
& TTC (min)

&0.30	&0.19	&0.72&	0.42&	0.36 & 0.36 &	5.2&	3.7&	2.9&	2.6& 2.5&	30.6&	25.5&	18.0	&15.3&	165.4&	165.4&	103.4&	82.7

\\[0.7ex] 
 & ITC

& 4663&	2157	&6545	&3704	&3127 & 3127&	15758&	11079&	8578	&7702&7691	&18000	&14828&	10416&	8828&	21481&	21326&	13270&	10585

\\[0.7ex] 
& Speedup

& 1$\times$
& \best{2.1$\times$} 

& 1$\times$
& \worst{1.7$\times$}
& \best{2$\times$}
& \best{2$\times$}

& 1$\times$
& \worst{1.4$\times$}  
& 1.8$\times$        
& \best{2$\times$}   
& \best{2.1$\times$}   

& 1$\times$
& \worst{1.2$\times$}   
& 1.7$\times$   
& \best{2$\times$}   

& 1$\times$
& \worst{1$\times$}   
& 1.6$\times$   
& \best{2$\times$}   
\\[0.7ex]

%% file: tables/cylinderalpha.tex
\multirow{3}{*}{ \makecell[l]{Cylinder\\$\alpha=1^{\circ}$}}
& TTC (min)

&0.36	&0.16&	0.68	&0.40&	0.34& 0.34	&5.0&	3.6	&2.9&	2.5& 2.5&	27.1&	22.6&	16.9&	13.6&	153.0&	139.1&	95.6&	76.5

\\[0.7ex] 
 & ITC

&4196	&1844&	6182	&3490	&2945 & 2941 &15152	&10646	&8736	&7399& 7400 	&15941	&13112	&9791	&7799	&19870	&17909	&12264	&9780
\\[0.7ex] 
& Speedup

& 1$\times$
& \best{2.2$\times$}  

& 1$\times$
& \worst{1.7$\times$}  
& \best{2$\times$}    
& \best{2$\times$}    

& 1$\times$
& \worst{1.4$\times$}  
& 1.7$\times$        
& \best{2$\times$}   
& \best{2$\times$}   

& 1$\times$
& \worst{1.2$\times$}   
& 1.6$\times$   
& \best{2$\times$}   

& 1$\times$
& \worst{1.1$\times$}   
& 1.6$\times$   
& \best{2$\times$}   
\\[0.7ex]

%% file: tables/TableWarmup.tex
\newcommand*\TableWarmup{
    \begin{table}[htbp]
	\centering
\scriptsize
\colorbox{lightgray}{
\setlength{\tabcolsep}{2.2pt}
\renewcommand{\arraystretch}{0.8}
	\begin{tabular}{l|l|cc|cc|cc|cc|cc}

		    \multicolumn{2}{l}{}
		    & \multicolumn{2}{c}{\makecell[c]{$64\times$ \\$256$ }}
		    & \multicolumn{2}{c}{\makecell[c]{$256\times$ \\$256$ }}
		    & \multicolumn{2}{c}{\makecell[c]{$512\times$ \\$512$ }}
		    & \multicolumn{2}{c}{\makecell[c]{$1024\times$ \\$1024$ }}
		    & \multicolumn{2}{c}{\makecell[c]{$2048\times$ \\$2048$ }}\\[0.7ex]
		    \cline{1-12}
		    \multicolumn{2}{c}{}&&&&&&&&&\\[-0.7em]

		    \multicolumn{2}{l}{Test case}
		    & W
		    & I
		    & W
		    & I
		    & W
		    & I
		    & W
		    & I
		    & W
		    & I\\[0.7ex]
\hline
		    &&&&&&&&&&&\\[-0.7em]
\multirow{2}{*}{ \makecell[l]{NACA 1412\\$\theta=5^{\circ}$}}
& T
		&2e-3 &3e-3 & 3e-3 & 1.3e-2 &  1.5e-2 & 0.05 & 0.11 & 0.25 & 0.77 & 1 \\[0.7ex]
& NI   
		&25 & & 26 & & 45  & & 65 & & 100& \\[0.7ex]
\hline
		    &&&&&&&&&&&\\[-0.7em]
\multirow{2}{*}{ \makecell[l]{NACA 1412\\$\alpha=6^{\circ}$}}
& T
		&2e-3 &3e-3 & 3e-3 & 1.3e-2 &  1.6e-2 & 0.05 & 0.13 & 0.25 & 0.85 & 1 \\[0.7ex]
& NI   
		&26 & & 26 & & 49  & &78 & & 110& \\
\hline
		    &&&&&&&&&&&\\[-0.7em]
\multirow{2}{*}{ \makecell[l]{NACA 0015\\$\theta=3^{\circ}$}}
& T
		&2e-3 &3e-3 & 3e-3 & 1.3e-2 &  1.7e-2 & 0.05 & 0.122 & 0.25 & 0.7 & 1 \\[0.7ex]
& NI   
		&25 & & 27 & & 50 & &72 & & 91& \\
		\hline
		    &&&&&&&&&&&\\[-0.7em]
\multirow{2}{*}{ \makecell[l]{NACA 0015\\$\alpha=0^{\circ}$}}
& T
		&3e-3 &3e-3 & 4e-3 & 1.3e-2 &  2e-2 & 0.05 & 0.14 & 0.25 & 0.92 & 1 \\[0.7ex]
& NI   
		&31 & & 39 & & 60 & &80 & & 119& \\
		\hline
		    &&&&&&&&&&&\\[-0.7em]
\multirow{2}{*}{ \makecell[l]{Ellipse\\$\theta=5^{\circ}$}}
& T
		&3e-3 &3e-3 & 4e-3 & 1.3e-2 &  2e-2 & 0.05 & 0.14 & 0.25 & 0.85 & 1 \\[0.7ex]
& NI   
		&35 & & 38 & & 69 & &80 & & 110& \\
		\hline
		    &&&&&&&&&&&\\[-0.7em]
\multirow{2}{*}{ \makecell[l]{Ellipse\\$\alpha=1^{\circ}$}}
& T
		&2e-3 &3e-3 & 4e-3 & 1.3e-2 &  1.8e-2 & 0.05 & 0.12 & 0.25 & 0.76 & 1 \\[0.7ex]
& NI   
		&28 & & 32 & & 55 & &73 & & 99& \\
		\hline
		    &&&&&&&&&&&\\[-0.7em]
\multirow{2}{*}{ \makecell[l]{Cylinder\\$\theta=0^{\circ}$}}
& T
		&3e-3 &3e-3 & 6e-3 & 1.3e-2 &  2.5e-2 & 0.05 & 0.14 & 0.25 & 1.05 & 1 \\[0.7ex]
& NI   
		&40 & & 50 & & 75 & & 84 & & 140& \\
		\hline
		    &&&&&&&&&&&\\[-0.7em]
\multirow{2}{*}{ \makecell[l]{Cylinder\\$\alpha=1^{\circ}$}}
& T
		&3e-3 &3e-3 & 6e-3 & 1.3e-2 &  2.5e-2 & 0.05 & 0.14 & 0.25 & 1.05 & 1 \\[0.7ex]
& NI   
		&42 & & 50 & & 76 & & 83 & & 142& \\
		\bottomrule
		\end{tabular}
	        \caption{\small Warmup (W) and inference (I) times (T) and number of iterations (NI) for each test case at each spatial resolution. Times are reported in minutes.\label{tab:warmup}}
		}
\end{table}
}

%% file: tables/TableBaseline.tex
\newcommand*\TableBaseline{
    \begin{table}[htbp]
	\centering
\scriptsize
\setlength{\tabcolsep}{5pt}
\renewcommand{\arraystretch}{0.8}
	    \begin{tabular}{l|cc|cc}

		    & \multicolumn{2}{c|}{\textbf{SURFNet}}
		    & \multicolumn{2}{c}{\textbf{BM}}\\[0.7ex]

		    &&&&\\[-0.7em]
		    & \makecell[l]{$256\times 256$ }
		    & \makecell[l]{$512\times 512$ }
		    & \makecell[l]{$256\times 256$ }
		    & \makecell[l]{$512\times 512$ }
		    \\[0.7ex]
\hline
		    &&&&\\[-0.7em]
		    \makecell[l]{Data collection\\time for training}
		    &
		    0.33
		    &
		    0.33
		    & 
		    1
		    &
		    3.86\\[0.7ex]
		    &&&&\\[-0.7em]
		    \makecell[l]{Data collection\\time for TL}
		    
		    &
		    0.06
		    & 
		    0.41
		    &
		    -
		    & 
		    - \\[0.7ex]
		    &&&&\\[-0.7em]
		    \makecell[l]{Training time}
		    &
		    2.5
		    & 
		    2.5
		    &
		    9.25
		    &
		    38\\[0.7ex]

		    &&&&\\[-0.7em]
		    \makecell[l]{TL time}
		    &
		    0.06
		    & 
		    0.5
		    &
		    -
		    &
		    -
		    \\[0.7ex]

		    \bottomrule
  \end{tabular}
  \caption{\small Comparison with the baseline model (BM) on data collection and training for reaching similar accuracy at $256\times256$ and $512\times512$ spatial resolutions.
  \label{tab:baseline}}
\end{table}
}

%% file: text/related.tex
\vspace{-0.5em}
\section{Related Work}
\label{sec:related}
\textbf{\emph{DL for CFD}}. Several recent approaches aim to find DL-based accelerators
for turbulent flows with promising results. 
\citeauthor{maulik} predict the eddy viscosity field, but not other flow properties such as the velocity~\cite{maulik}.
\citeauthor{unet} use an encoder-decoder type of network but their approach does not account for the eddy viscosity field~\cite{unet}. 
Although they present real-time solutions, the geometry in the training
and prediction stages are same (\ie airfoils) making the solution less
generalizable.
Alternatively, ~\citeauthor{cfdnet} feed the network's prediction back into the
physics solver to enable generalization with the same model without relaxing the convergence constraints. 
These efforts are not resolution-invariant unless the network is trained with large-scale data from a variety of high-resolution simulations. 
~\citeauthor{tompson} accelerate Eulerian fluid simulations by minimizing the divergence of the velocity field using an unsupervised method. 
This approach~\cite{SC19} maintains accuracy up to $1024\times 1024$ resolution. 
However, Eulerian fluid simulations ignore the viscous effects that are critical for most engineering systems of interest. 
SURFNet predicts all relevant fluid variables in the entire domain of turbulent flows to accelerate CFD simulations.

\textbf{\emph{Mesh-independent DL approaches}}.~\citeauthor{pinns} introduced \emph{physics-informed neural networks} (PINNs) - networks trained to respect any physics laws.  
These methods substitute traditional solvers~\cite{pinns,sciann,deepxde}. 
However, this approach has several constraints. 
First, for complex turbulent flow problems, including the conservation laws in the loss function may lead to stiffer optimizations~\cite{pinnstiff}. 
Second, training a new model is required for every new instance of a distinct flow configuration (unless the initial/boundary conditions are an input to the network~\cite{trainonceuseforever}), instead of a simple forward pass of the network, severely restricting its generalization capabilities. 
~\citeauthor{deeponet} introduced an infinite-dimensional operator with neural
networks, known as \emph{neural operators} (NO), that learns the nonlinear operation from partial differential equations without knowledge of the underlying PDE -- only with data. 
NO provides a single set of network parameters that are compatible with different discretizations.
Hence, they are resolution-invariant.
However, these approaches~\cite{fourieroperator,graphkernelnetwork,bhattacharya} train the network with data downsampled from high-resolution simulations -- which is impractical for many practitioners and suffers from the same computational constraints of traditional solvers. ~\citeauthor{meshfreeflownet} introduced a CNN-based resolution-invariant approach that satisfies the underlying PDE. However, it also suffers from the same data-collection limitation as the former approaches.
SURFNet overcomes the above limitations by training the CNN models primarily from data gathered at low-resolutions while enabling super-resolution by only minimal fine-tuning. 

%% file: text/conclusion.tex
\vspace{-0.75em}
\section{Conclusions}
This paper presented SURFNet, a super-resolution flow network that accelerates high-resolution turbulent CFD simulations.
SURFNet is primarily trained on low-resolution simulation data and applies this information (via transfer learning) to high-resolution inputs.
We proposed two variations, one-shot transfer learning (OSTL) and incremental transfer learning (ITL). 
Both approaches yield consistent speedups across test geometries unseen during the training or transfer stages and exhibit good generalization capacities. 
We demonstrated resolution-invariance with ITL on domains up to 256$\times$ larger than the tiny discretization used in training and a uniform $2-2.1\times$ speedup across target resolutions and test geometries compared to the OpenFOAM CFD solver. SURFNet is able to recover high-resolution flow features with $15\times$ less data at high resolutions and reducing the combined data collection and training time by $3.6\times$ and $10.2\times$ at $256\times256$ and $512\times512$ grid sizes, respectively. 

Future work includes expanding SURFNet to a wide range of flow and boundary conditions.
As SURFNet is computationally inexpensive, and
data from low-resolution CFD simulations are widely available, further explorations of the generalization capabilities are warranted. 
Since SURFNet's underlying CNN architecture is
unconstrained, SURFNet could be beneficial for transfer learning across physical domains, such as molecular dynamics or solid mechanics.